\documentclass[letterpaper, twocolumn, 10pt, table]{article}
\usepackage{usenix-2020-09}

\usepackage{booktabs}
\usepackage{changepage}
\usepackage{graphicx}
\usepackage{hyperref}
\usepackage{longtable}
\usepackage{soul}
\usepackage{subcaption}
\usepackage[disable]{todonotes}
\usepackage{todonotes}
\usepackage{xcolor}
\usepackage{xspace}

\def\rtimeloops{Timeloops\xspace}
\def\timeloops{\textsc{\rtimeloops}\xspace}
\def\ptitle{\rtimeloops: Automated System Call Policy Learning for
	Containerized Microservices}
\def\pkeywords{} %
\def\eg{e.g.,\xspace}
\def\ie{i.e.,\xspace}
\def\etc{etc.\xspace}

\graphicspath{ {./figures/} }

\hypersetup{
	colorlinks	= true,
	citecolor	= {red!70!black},
	filecolor	= black,
	linkcolor	= {green!70!black},
	urlcolor	= {blue!70!black},
	pagebackref	= false,
	pdftitle	= {\ptitle},
	pdfkeywords	= {\pkeywords}
}

\def\addauthnote#1#2{
	\expandafter\def\csname#1\endcsname##1{\todo[inline,size=\footnotesize,color=#2]{\textbf{\underline{\texttt{#1}}:} ##1}\xspace}
}

\addauthnote{vpk}{red!25}
\addauthnote{meghna}{blue!25}
\addauthnote{andreas}{green!25}
\addauthnote{simha}{yellow!25}

\definecolor{lightgray}{gray}{0.9}

\begin{document}

\date{}

\title{\Large \bf \ptitle}
\author{
{\rm Meghna Pancholi}\\
meghna@cs.columbia.edu\\
Columbia University
\and
{\rm Andreas D. Kellas}\\
andreas.kellas@cs.columbia.edu\\
Columbia University
\and
{\rm Vasileios P. Kemerlis}\\
vpk@cs.brown.edu\\
Brown University
\and
{\rm Simha Sethumadhavan}\\
simha@columbia.edu\\
Columbia University
} %
\maketitle

\begin{abstract}

We introduce \timeloops, a novel technique for automatically learning system
call filtering policies for containerized microservices applications. At
run-time, \timeloops automatically learns which system calls a program should
be allowed to invoke, while rejecting attempts to call spurious system calls.
Further, \timeloops addresses many of the shortcomings of state-of-the-art
static analysis-based techniques, such as the ability to generate tight filters
for programs written in interpreted languages such as PHP, Python, and
JavaScript. \timeloops has a simple and robust implementation because it is
mainly built out of commodity, and proven, technologies such as seccomp-BPF,
\texttt{systemd}, and Podman containers.
We demonstrate the utility of \timeloops by learning system calls for
individual services and two microservices benchmark applications, which utilize
popular technologies like Python Flask, Nginx (with PHP and Lua modules),
Apache Thrift, Memcached, Redis, and MongoDB. Further, the amortized
performance of \timeloops~ is similar to that of an unhardened system, while
producing a smaller system call filter than state-of-the-art static
analysis-based techniques.

\end{abstract}

\section{Introduction}
\label{sec:introduction}

Microservices have become prominent due to shifts from monolithic to
modularized and distributed architectures. While monoliths are complex and
large services running on a single host, microservices are lightweight,
loosely-coupled services running in a distributed fashion. Inter-service
network communication allows these small components, which are typically
containerized, to collaborate and complete actions that would previously be
completed by a single monolith. Microservices have become a popular choice due
to their modularity, which simplifies the development of large applications,
allows for elastic scaling, and supports language and framework heterogeneity.
More importantly, microservice deployments are quickly rising in popularity:
surveys conducted by O'Reilly in 2020 concluded that $77\%$ of tech companies
have adopted this model for computing~\cite{loukidesswoyer2020}.

Microservice deployments, however, introduce two main complexities that make
security challenging: (1)~{\bf Dynamism:} graphs of microservices architectures
evolve rapidly due to new application features~\cite{dsb}, making it difficult
to keep security policies up to date; (2)~{\bf Heterogeneity:} each service in
an application can be developed by different teams in dissimilar programming
languages, and hosted on completely different platforms, sometimes across data
centers, increasing the attack surface. These challenges are also compounded to
a degree by the need to provide low end-to-end latency. In the absence of good
security protection for microservices, there is risk not only of application
compromises but also a risk to other tenants via container escape
attacks~\cite{li2021automatic}.

State-of-the art efforts for securing a microservice in industry and academia
focus on two main techniques for protecting microservices: (a)~{\bf Rootless
containers:} increasingly containers offer rootless deployment options that
permit containers to be run as unprivileged users---this raises the bar to gain
privileged execution on a system via a compromised microservice; (b)~{\bf
System call filtering:} even with rootless containers, compromised
microservices can be used to gain privileges via the system call API. As such,
system call filtering is used to confine the process' ability to make arbitrary
system calls, or invoke buggy ones~\cite{li2017lock}, thus mitigating this
risk~\cite{ghavamnia2020confine, kim2021prof, host-security, priv-esc}. An
added benefit of system call filtering is that it does not significantly impact
the latency of the microservice.

In this paper, we propose a new method for securing microservices called
\timeloops. We designed \timeloops with microservices' strengths and weaknesses
in mind. For each service in a microservices application, the \timeloops
technique relies on three components for learning legitimate system calls:
(1)~a production service, (2)~an oracle service, and (3)~a \timeloops
controller. The production service is an unhardened service that is tuned for
low latency, while the oracle service is a hardened service that is tuned for
security. An example of an oracle service could be a microservice running with
compiler-inserted, runtime checks to protect against memory safety errors.
The high level idea is to sparingly use the oracle service to decide if a
system call should be added to the allow list. This provides the security
guarantees of a hardened microservice without constantly incurring the
overheads associated with hardening.

Compared to state-of-the-art microservice security techniques, our solution has
two unique advantages: (i)~\timeloops is language agnostic, since microservice
deployments tend to be heterogeneous with services written in different
languages. Traditional system call filtering approaches that use static
analysis to create filters~\cite{temp-sys-filtering, canella2021automating}
require writing custom tools for each language. This increases the difficulty
of applying these methodologies to modern cloud microservices---\timeloops
avoids this problem completely. (ii)~\timeloops offers a tighter set of system
call constraints that is customized to each service. Unlike static analysis
systems, \timeloops learns the system calls required per service using runtime
workloads and therefore produces a much tighter set.

Using these components we measure both the security and performance of our
\timeloops system. We observe that the amortized performance of a
\timeloops-protected system is similar to that of an unhardened, insecure
system, and our approach nearly eliminates all the performance overheads
associated with hardening. In terms of the tightness of the system call
filtering set, we show that a state-of-the-art static analysis tool generated
system call sets that were $32.7\%$ larger than the ones generated by
\timeloops.  Additionally, we demonstrate that \timeloops succeeds in creating
system call filters for interpreted languages with very little effort, which is
a challenge for existing static analysis-based techniques. Finally, as a side
effect, we show that we can easily detect memory corruption-based exploits
while taking place.

The remainder of this paper is organized as follows. In
Section~\ref{sec:background}, we provide the necessary background information
and present our threat model. In Section~\ref{sec:applied}, we introduce
\timeloops, a novel approach for learning new security policies, which
leverages runtime exploit detection techniques, while providing amortized
runtime performance overhead. We consider the security architecture of
\timeloops in Section~\ref{sec:security}, and evaluate our prototype's runtime
performance and filter learning correctness in Section~\ref{sec:performance}. We
discuss related work in Section~\ref{sec:relatedwork}, and conclude our work in
Section~\ref{sec:conclusion}.

\section{Background and Threat Model}
\label{sec:background}

\begin{figure}[!t]
  \centering
  \includegraphics[width=\linewidth]{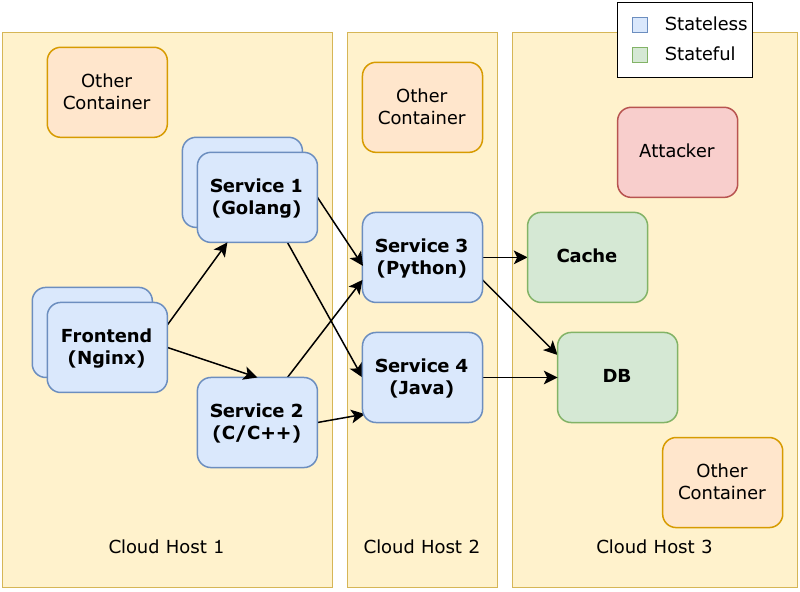}
  \caption{Microservices applications are the composite of diverse containers
      communicating to perform a task. They are scheduled to run on various
      multitenant hosts with other potentially malicious containers. Most of
      these services are stateless and are scaled and restarted on demand.}
  \label{fig:microservices-landscape}
\end{figure}

\subsection{Microservices Overview}

{\bf Microservices\xspace}
Microservices have become a de facto standard for building and deploying
web-based services. Building software as a set of communicating lightweight
processes allows multiple teams to work independently on small individual units
of code---\ie~microservices---that can then be combined together using only API
calls exposed by the service~\cite{edprice-msft}. For example, if we consider
the set of services running in Figure~\ref{fig:microservices-landscape}, a team
may be responsible for maintaining Service~1, independently, while another team
may manage Services~3 and 4 without needing to understand the implementation
details of the other services. The development of microservice-based
applications has been made easier by the use of popular languages and
libraries, like Python, Thrift, and Node.js. Microservice environment
management has become configurable and reproducible through container
technologies, like Docker and Podman. Cloud platforms like Amazon EC2, Google
Cloud Platform, and Microsoft Azure have made deployments easy to schedule and
dynamically scale.

Since microservices are typically deployed in multitenant cloud environments,
they have to be written in a manner that makes them resilient to crashes and
long-latency inducing events such as network congestion. A very common design
pattern is to write these microservices in a stateless manner with retry
semantics \ie, in a manner that does not preserve state across requests and
that produces correct results even when multiple repeated queries are issued by
a client. In fact, most cloud deployments today isolate the stateless frontend,
routing, or computation services from the stateful database or caching
services, where service crashes and restarts do not affect the correct behavior
of the (composite) system. To ensure successful execution of their requests
despite frequent restarts and long-latency events, clients will often resend
identical, idempotent requests until the requests are satisfied.

While the modularity of microservices has enabled many desirable software
engineering practices, like rapid feature development, and continuous
integration/\-delivery (CI/CD), it has also created some serious security
challenges. It is not uncommon today to have deployed applications with
hundreds or even thousands of microservices~\cite{cockcroftms}, written in many
different languages, such as Python, PHP, Javascript, C/C++, Rust, Go, \etc
Frequent changes to these software pieces rules out manually writing security
policies because expected behavior is hard to define in such a constantly
changing environment; on the other hand, code analysis techniques to understand
application behavior are less effective in the presence of excess library code
and with interpreted languages such as Python and Javascript. Besides, multiple
microservice applications can run in a multitenant manner, on a single host, on
a cloud with no guarantees about the services running alongside any given
container. Container escape vulnerabilities and lateral movement attacks pose
huge threats to deployments of this style \cite{chierici2022, ren_liu_2016}.

{\bf Exploitation and Mitigations\xspace}
When attackers attempt to execute code in a container's application---\ie~a
microservice,---they are limited by the isolation boundaries established by the
container runtime such as namespaces or control groups
(\texttt{cgroups})~\cite{rice2020}. However, attackers can sometimes ``escape''
from the container's environment, and this allows them to potentially access
system resources~\cite{containerescapology} or escalate their privileges.

In order for an attacker to gain arbitrary code execution and execute system
calls for nefarious purposes, in reasonably secure microservices, they often
must exploit a vulnerability in the microservice. To prevent exploitation, a
wide range of research proposals that aim at hardening application code, using
runtime verification and enforcement mechanisms~\cite{szekeres2013sok,
larsen2014sok, burow2019sok, song2019sok}, can also be applied to
microservices. For example, SoftBound~\cite{nagarakatte2009softbound} and
CETS~\cite{nagarakatte2010cets}, as well as AddressSanitizer
(ASan)~\cite{serebryany2012addresssanitizer}, provide detection of
memory-safety-related issues (\eg~out-of-bounds accesses and uses of freed
objects). However, these software techniques have fairly high overhead ranging
from $70\%$ to $300\%$ depending on the level of protection. Due to these high
overheads, many of them are usually employed in security auditing and testing
environments and not in production deployments, which tend to be latency
sensitive---more so for microservices that rely on expensive network calls for
communication.

Because runtime vulnerability mitigation cannot be used for microservices, to
prevent container escapes, low overhead mitigation techniques are typically
used. A very popular method is to use \verb$seccomp-BPF$ to restrict the system
calls available to a container application~\cite{ghavamnia2020confine,
kim2021prof}. Common container implementations (\eg Docker, Podman) allow easy
installation of \verb$seccomp-BPF$ filters and even provide a default filter
that removes access to $44$ system calls for all containers. The container
\verb$seccomp-BPF$ interface can be leveraged to customize and further restrict
the system calls available to an microservice executing in a
container~\cite{dockerseccomp}.

\subsection{System Call Filtering}

The system call (syscall) interface enables user-space applications to request
privileged services from OS kernel. This interface is quite large: for example,
the Linux kernel (v5.x) provides $\approx$$350$ syscalls to user-space
applications~\cite{sysfilter}. Like most programs, microservices only need a
subset of these syscalls for proper execution~\cite{sysfilter,
ghavamnia2020confine, temp-sys-filtering, canella2021automating, kim2021prof,
chestnut}. An attacker, however, who is able to leverage a vulnerability in a
microservice to gain arbitrary code execution can use \emph{any} of the
available syscalls, effectively (1)~violating the principle of least privilege
and (2)~further (ab)using vulnerabilities in less-stressed kernel code paths to
escalate privilege~\cite{li2017lock}. By restricting the system calls available
to a microservice, an attacker is constrained to only performing actions that
fall within the benign behavior(s) of the victim program.

Determining which system calls should be allowed for a microservice is
challenging. Policies can be created manually, but with significant developer
effort~\cite{dockerseccomp}. Therefore, attempts to synthesize policies
automatically have become more prevalent~\cite{sysfilter, ghavamnia2020confine,
temp-sys-filtering, canella2021automating, kim2021prof}. Current approaches for
automatically generating system call policies oftentimes use either static or
dynamic analyses (or a combination thereof). Static analyses attempt to reason
about the system calls that a target program should be allowed to execute
without executing the program~\cite{sysfilter, ghavamnia2020confine,
temp-sys-filtering}. These approaches are generally \textit{over-approximate}
in their analyses because they explore all possible code paths a program may
execute. Dynamic analyses typically execute the target program with some set of
known-safe training inputs, like developer-written tests, in order to record
all executed system calls and to populate an allow-list from
them~\cite{canella2021automating, kim2021prof}. These approaches are generally
\textit{under-approximate}, because they are only able to identify system calls
that are observed during execution of training data, and fail when a new
legitimate system call is executed.

For both static and dynamic system call policy extraction, current approaches
for syscall extraction and enforcement create \emph{immutable} policies that
cannot be later refined. In an immutable system, when the target program
attempts to execute a non-allowed system call, the enforcing mechanism will
terminate the program in order to prevent potential exploitation. This occurs
at the risk of the system call being the result of a safe input that was
incorrectly omitted from the training data---\ie~a false positive. In order for
it to be flexible enough to update its policy once deployed, the system must be
able to distinguish the execution of syscalls that result from benign and
offending inputs at runtime.

Once a system call policy is created in the form of an allow- or deny-list, it
must be enforced. The Linux kernel provides the `SECure COMPuting'
(\verb$seccomp$) mode that allows users to restrict the system calls that a
process is allowed to make~\cite{manseccomp}. With custom BSD Packet Filters
(BPF)~\cite{mccanne1993bsd}, \verb$seccomp-BPF$ allows a user to configure the
behavior of the kernel when the user-space process makes a non-allowed system
call; for example, the kernel can be configured to send a \verb$SIGKILL$ or
\verb$SIGSYS$ signal to the process, depending on the arguments provided via
the \verb$seccomp-BPF$ interface, log the event to an audit log,
\etc~\cite{seccomp}.

\subsection{Threat Model}
\label{sec:threatmodel}

In this work, we consider an attacker who is able to gain remote code execution
in victim network applications. That is, the attacker can provide inputs to a
program over a network connection to exploit vulnerabilities in the program and
introduce new, unintended code or behavior in the victim process. Our model
assumes that the attacker does not have physical access to the host machine and
considers side-channel~\cite{spectre, meltdown} and fault~\cite{clkscrew,
plundervolt} attacks out-of-scope.

In our model, the victim application executes in a container environment that
is isolated from the host system resources, so we consider that the attacker is
motivated to ``escape'' from the container by using remote code execution to
remove the isolation between the victim application and the host resources. The
attacker is able to exploit some vulnerability or misconfiguration to execute
code in the container process~\cite{szekeres2013sok}, and may desire to escape
the container environment or compromise the container application for malicious
purposes \emph{via executing arbitrary system calls}~\cite{containerescapology,
abubakar2021shard, li2017lock}. Our threat model is in par with prior work in
the area~\cite{sysfilter, temp-sys-filtering, canella2021automating, chestnut}.

\section{Approach Overview}
\label{sec:design}

For a system to be able to learn a security policy at runtime, it must be able
to determine whether a violation of the current policy is the result of a
benign input that was previously unseen or the result of an offending input
that should be prevented by the policy. In this section, we describe a novel
approach for learning software security policies by introducing an
\textit{oracle} to assist with determinations of safe inputs.

Our policy learning approach provides a mechanism to learn from policy
violations. When a violation occurs after the program executes a specific
input, we consult the oracle service by providing it with the current policy
and input, and re-execute the program under the oracle's observation. The
oracle makes a determination that the security policy violation occurred either
because of a benign input to the program or because of offending input to the
program. If the oracle determines that the policy violation was due to a benign
input, then the policy is updated to allow the event that was observed, and the
program can be restarted with the updated policy. If the oracle determines that
the policy violation was due to offending input, then an exploit attempt was
prevented and an alert is raised accordingly.

For this approach to succeed, the program must be resilient to repeated
re-executions from a known state, whether from the program start or from some
checkpoint location. We observe that network services, and in particular,
microservices, are often designed to be stateless and support the re-execution
of idempotent request operations. Because of this, we can take advantage of
\textit{retry semantics} or we can record and replay requests without affecting
the correctness of the entire application's state. For engineering simplicity,
we assume that retry semantics exist and our application's clients resend
requests until they succeed. These types of services can be re-executed and are
thus good candidates for a timelooping security policy learning approach.

\section{Design and Implementation}
\label{sec:applied}

\begin{figure}[t!]
  \centering
  \includegraphics[width=\linewidth]{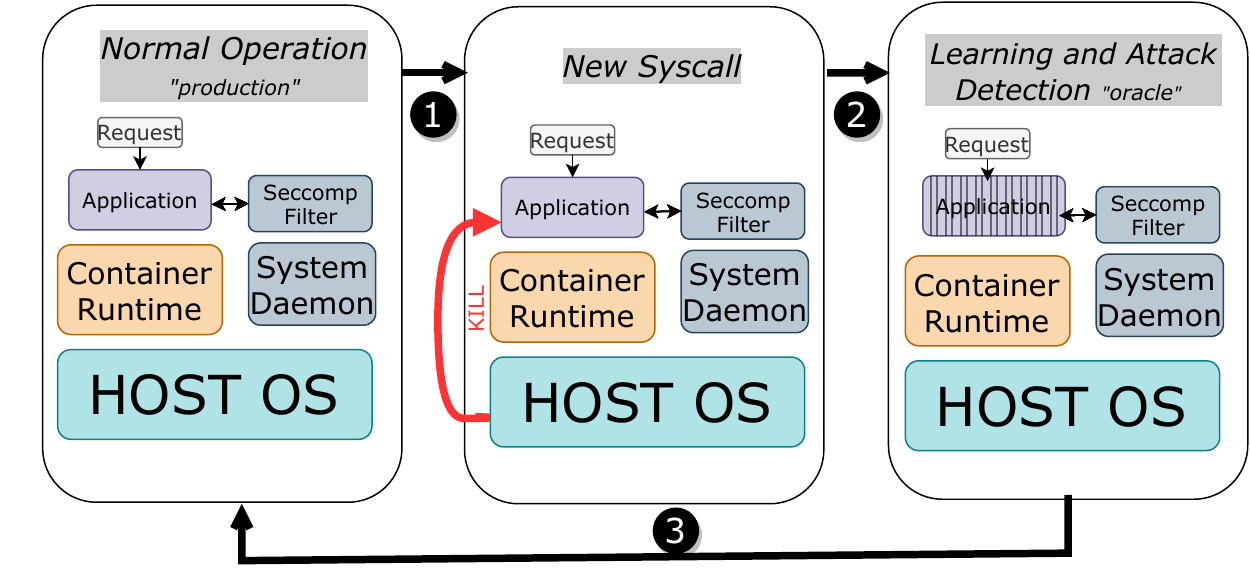}
  \caption{The \timeloops controller initializes by starting the production
      version of the service container, which services incoming requests. When a
      system call violation occurs, (1)~the service container is killed by a
      signal sent from the host OS, and then (2)~the \timeloops
      controller starts the oracle container. The oracle container executes a
      hardened version of the service, which receives the next incoming request
      and determines whether it is a safe or offending input. The oracle makes a
      determination and exits, and (3)~the \timeloops controller uses the
      oracle's determination to decide how to update the policy. The controller
      starts the service container with the most recently updated system call
      policy.}
  \label{fig:flow}
\end{figure}

In this section, we describe the design and implementation of the \timeloops
system, which learns systems call policies on-the-fly, as the programs execute,
thereby including only system calls that are actually observed and detecting
malicious behavior when a new system call is introduced. The sequence of events
that occur when a service running in \timeloops receives an input that causes a
system call violation is illustrated in Figure \ref{fig:flow}.

\subsection{System Call Policy Learning}
\label{sec:applied-design}

The \timeloops~system is designed as three main components: (1) the service,
(2) the oracle, and (3) the controller. Both the service and oracle are
deployed in their own isolated container environments, while the controller is
responsible for managing the execution of the service and oracle containers and
for applying policy changes as information is learned.

The \timeloops~controller begins execution by starting the service container
with some default policy of allowed syscalls (which may be initialized as an
empty list). If, given some input, the service container attempts to execute a
syscall that is not in the allow-list, the service is forced to exit and the
controller is notified. The controller then starts the oracle container, which,
in turn, starts the same service but with additional runtime exploit
mitigations deployed; for example, the service can be compiled with ASan to act
as an oracle for some types of memory corruption-based exploits. Importantly,
the controller starts the oracle service \textit{with the same allow-list of
system calls that was applied to the service}. Given the same input, the oracle
should either (1) exit due to the same system call violation, or (2) exit due
to an ASan abort. Case (1) indicates that the code path taken to reach the
system call did \textit{not} result in memory corruption, and so the offending
system call may safely be added to the allow-list---in this case, the
controller updates the policy and starts the original service with the new
policy. Case (2) indicates that prior to reaching the new system call, a memory
corruption violation occurred, and that the input is unsafe.

To summarize, \timeloops~identifies when a new system call violates an existing
policy, queries an oracle to determine whether the input that caused the system
call execution is ``safe'' according to the runtime exploit detection features
of the oracle, and makes policy update decisions based on the determination. An
additional benefit is that the runtime performance overhead of executing a
program with exploit detection and mitigations applied occurs \textit{lazily},
\ie~only when a system call policy violation occurs. This means that in the
critical path of the program execution, the overall system executes with the
amortized runtime of the service, while still providing the benefits of
executing exploit detection checks opportunistically. We study and analyze the
runtime performance of \timeloops~in Section~\ref{sec:evaluation}.

\subsection{System Call Policy Enforcement}

The service and oracle programs are isolated in their own container
environments, as we would see in typical microservices deployments. This
provides separation from system resources and from the controller service that
manages the security policy. We use Podman containers \cite{podman} in the
\timeloops~implementation, which is an alternative to Docker, the most popular
container engine. Podman's security advantage is its daemonless
architecture---unlike Docker, which uses a single daemon executing with
administrative privileges to manage containers, Podman can create containers as
non-\texttt{root} child processes. Podman is compliant with the Open-Container
Initiative (OCI) standard, which makes it easy to use and compatible with
existing Docker images. The \timeloops~controller leverages these features to
start and stop the service, and oracle containers as \verb$systemd$ services.

Both Podman and other container runtimes provide support for applying a
\verb$seccomp-BPF$ filter to a container on start-up. This functionality is
enabled by the default filter~\cite{dockerdocs} and removes access to at least
$44$ system calls that can be abused for container escape exploits, but which
still leaves over $300$ system calls available. We use the Podman feature to
apply a custom \verb$seccomp-BPF$ filter every time the \timeloops~controller
starts the service and oracle containers. In between executions of the
containers, the controller updates the filter, but once a container is started
with a given filter, the set of syscalls that the container is allowed to
execute is immutable until the container exits (by the \verb$seccomp-BPF$
design).

\subsection{System Calls Introduced by the Oracle}

It is possible for the oracle container to execute system calls that the
original service did not. This occurs when the oracle's exploit detection
mechanism introduces new system calls or non-determinstic behavior takes place.
To handle this, we instruct our \timeloops controller to add \textit{any}
system call that the oracle encounters during execution to the allow-list,
assuming that the oracle does not detect exploitation prior to the system call
being executed. These system calls can be trusted because the oracle is the
determinant of safe inputs. However, introducing these syscalls potentially
causes our allow-list to be looser than necessary and requires extra iterations
of \timeloops learning, so ideally we want to choose a hardening technique that
provides strong security guarantees without introducing many new system calls.

\subsection{Retry-request Expectations}

In the current \timeloops implementation, the service may receive an input that
results in a system call violation, causing the service to exit without
handling the request. \timeloops then restarts the service container in the
oracle and awaits the next request. We expect the client to use retry
semantics, meaning that the client retries the request until successful. These
expectations also align with the behavior of clients interacting with
idempotent microservices often expect retry semantics until a response confirms
success for robustness and fault tolerance~\cite{tail-at-scale}. Additionally,
the expectation that identical messages are repeated does not introduce
security vulnerabilities in the \timeloops~system when the expectation is not
met---it only makes the system inefficient. The security implications of this
are discussed further in Section~\ref{sec:security}.

\subsection{Limitations of Policy Learning}

In our implementation, we use ASan as the runtime exploit detection technique
for the oracle since it provides instrumentation for detecting runtime memory
safety violations. To create an oracle container for a given service, we use
Clang to compile the service's code and dependencies with ASan. Our flexible
design allows users to create oracle containers hardened with other exploit
detection techniques, customized to their threat landscape. We opt for ASan
since $\approx$$70\%$ of CVEs each year are attributed to memory safety
issues~\cite{team2019}.

Our system determines when a new system call is the result of a memory
corruption exploit that ASan detects. However, when a system call is introduced
by an attacker via an exploit or misconfiguration that does \textit{not}
leverage a memory corruption vulnerability, \timeloops~will update the system
call policy to allow the new system call. Therefore, the scope of the security
policies that \timeloops~can learn is limited by the abilities of the oracle.
In Section~\ref{sec:attack-categories}, we discuss complimentary approaches for
maintaining the security of the service container and the overall
\timeloops~system in the context of exploits beyond what the oracle can detect.
We also intend for the \timeloops~oracle model to be modular, such that
multiple oracle containers with different forms of analyses can be deployed in
parallel to alleviate some of the pressure of finding one perfect hardening
mechanism.

\section{Evaluation Testbed}
\label{sec:evaluation}

We conduct our analyses by executing four different applications in \timeloops.
Two applications are statically served by Python Flask and Nginx, and two
applications are the social network and media microservices benchmarks from the
DeathStarBench suite~\cite{dsb}.

\subsection{Python Flask}

Flask is a popular Python web development framework used in microservices. We
developed a simple Flask application that serves a static website over HTTP(S)
to evaluate our \timeloops prototype. Flask applications are executed by the
Python interpreter and provide multi-threaded functionality. Many web
applications are developed using similar frameworks and interpreted languages,
like PHP, Ruby, and JavaScript. Since Python and its modules facilitate
developing web applications, or querying databases, easily, it is useful for
the rapid development of microservices, and thus has become a popular choice
for cloud developers and a suitable application to evaluate
\timeloops~\cite{hedgeskeating2021}.

\subsection{Nginx and PHP}

Nginx is a free and open-source web server that is used by $32.9\%$ of all
websites, the largest of any available web server~\cite{w3techs-nginx}. It is
often used as a gateway or load-balancer to multi-tier microservices
applications. We built a simple, static website hosted by an Nginx server with
PHP-FPM integrations---PHP is used by $78.1\%$ of websites where the
server-side language is known~\cite{w3techs-php}. Both Nginx and PHP-FPM use a
forking server model that spawns worker processes to handle connections. We
evaluate \timeloops~with this application to represent services that use
multiple processes to provide web content. Additionally, the PHP modules used
in this Nginx application rely on an interpreter which makes predicting its
behavior difficult, but suitable for our application like the Python Flask
application.

\subsection{DeathStarBench Microservices}

In order to evaluate \timeloops in an environment representative of real-world
microservices applications, we selected applications from the DeathStarBench
benchmark suite of microservices~\cite{dsb}. We specifically evaluated against
the media streaming benchmark, which models the structure of popular media
streaming services like Netflix, and the social network benchmark, which models
a social network site like Twitter. Both applications depend on technologies
used in real microservice deployments such as Memcached, Redis, Thrift,
MongoDB, and Nginx, and its services are written in a wide variety of
programming languages.

To deploy these services, we ran each each microservice from a benchmark on a
different Amazon EC2 instance~\cite{ec2}, similarly to how it might be deployed
in practice.  Each EC2 instance had its own \timeloops~controller setup and
generated a unique filter for each service.

To evaluate the social media and media streaming applications, we use the
workload generator provided by the benchmark suite. The workload generated for
evaluation resulted in the creation of many requests within the tiers of the
microservices application and required all services to undergo many iterations
of \timeloops policy learning. This way, we ensured that this workload is
suitable for not only performance benchmarking, but also our \timeloops
security evaluation.

\section{Security Evaluation}
\label{sec:security}

In this section, we consider the security features of \timeloops, and present
the argument that it provides protection to a container application given
reasonable assumptions about the threat model considered.

\subsection{Attacks Classification}
\label{sec:attack-categories}

\begin{table*}[h!]
  \begin{center}
    \textbf{Attacker Capabilities in \timeloops vs.
	Static System Call Filtering}
    \begin{tabular}{ |m{3cm}||m{5cm}|m{5cm}| }
      \hline
        & New System Calls Executed & No New System Calls Executed \\
      \hline\hline
      Oracle-Detectable Vulnerability & (1) \timeloops always detects the attack
      and stops execution, but static system call filtering can only do so if
      filter is adequately tight. & (2) The attacker is limited to using the
      existing system call filter. The filter is tighter with \timeloops since
      it only adds previously seen syscalls. \\
      \hline
      Oracle-Undetectable Vulnerability & (4) When \timeloops is supplemented
      with static-analysis-generated filters, \timeloops security is equal to
      that of static system call filtering. & (3) The attacker is limited to
      using the existing system call filter. The filter is tighter with
      \timeloops since it only adds previously seen syscalls. \\
      \hline
    \end{tabular}
  \end{center}
  \caption{\timeloops provides strong security benefits in comparison to
  previous works that rely on generating a static system call filter. Static
  system call filtering tools generate a filter prior to execution and only
  enforce that filter throughout application execution. Our tight and
  incremental filter building process results in a smaller attack surface and
  restricts attacker capabilities when a service is under attack.}
  \label{fig:sec-table}
\end{table*}

Given the presence of a remote attacker, as described in our threat model (see
Section~\ref{sec:threatmodel}), we classify attack scenarios by considering
whether the attacker invokes new system calls and whether the attacker
leverages oracle-detectable vulnerabilities. The security of \timeloops hinges
on detecting system call violations and detecting attacks with the oracle,
which is why we chose these components to rigorously analyze. These four attack
categories are shown in Table~\ref{fig:sec-table}, and in this section we
explain how the \timeloops~implementation provides varying security benefits in
each attack scenario.

Category (1) attacks (oracle-detectable exploit that executes a
policy-violation system call) are detected and prevented by the \timeloops
system, and \timeloops~provides the strongest security benefits against these
attacks. When the application container violates the system call policy as a
result of offending input, the oracle container will detect that the input is
an exploit trigger, prevent any changes to the security policy, and alert an
administrator.

Category (2) attacks (oracle-detectable exploit that does not execute a
policy-violation system call) and category (3) attacks (oracle-undetectable
exploit that does not execute a policy-violation system call) are handled in
the same way by the \timeloops. In these attack scenarios, \timeloops's
security benefits are in line with the approach of enforcing the Principle of
Least Privilege (PoPL)~\cite{saltzer1975protection}, the goal of previous work
in system call filtering. Because the system call policy is never violated in
Category (2) and (3) attacks, the oracle is never consulted, but an attacker is
still severely limited in terms of its capabilities with respect to accessing
the system call API. Due to the tight system call policy learned, the attacker
is restricted to only invoking previously observed system calls, thereby
limiting attempts to compromise the application or escape from the container.
Unlike previous works that create immutable lists that include all system calls
an application requires, \timeloops is a significant improvement because system
calls are conservatively added as they are observed. Hence, the \timeloops
allow-list can be much smaller than the set of all syscalls required by the
application.

Category (4) attacks (oracle-undetectable exploit that executes a
policy-violation system call) may seem the most dangerous to the
\timeloops~system, but with relatively-simple modifications, \timeloops~is able
to prevent the execution of arbitrary system calls. When an input from this
attack category is received, the oracle service is consulted to make a
determination on the input, but the oracle is unable to detect an exploit. This
grants attackers the ability to add any system calls to the allow-list until a
sufficient set are available to conduct an exploit. Therefore, to mitigate the
chance of this occurring, we propose two best practices when deploying an
application with \timeloops. First, an oracle must be carefully chosen
considering the application and the nature of its vulnerabilities. Second, we
can specify a list of system calls that are known to be unnecessary and common
in exploits and prevent them from ever being added to our system call policy.
For example, an over-approximate static analysis tool or the system calls in
the default Docker and Podman \verb$seccomp-BPF$ filter can be used to create
this list. For stronger guarantees, a static analysis-based technique for
generating system call filters can also be applied to create a list of system
calls that may never be included in the~\timeloops filter~\cite{sysfilter}.
With this mitigation, we see that~\timeloops security guarantees are equal to
those of static-analysis based techniques, but in many most cases, much
stronger.

\subsection{System Call Policies}\label{sec:evaluation-policies}

\begin{figure}[t]
    \includegraphics[width=\linewidth]{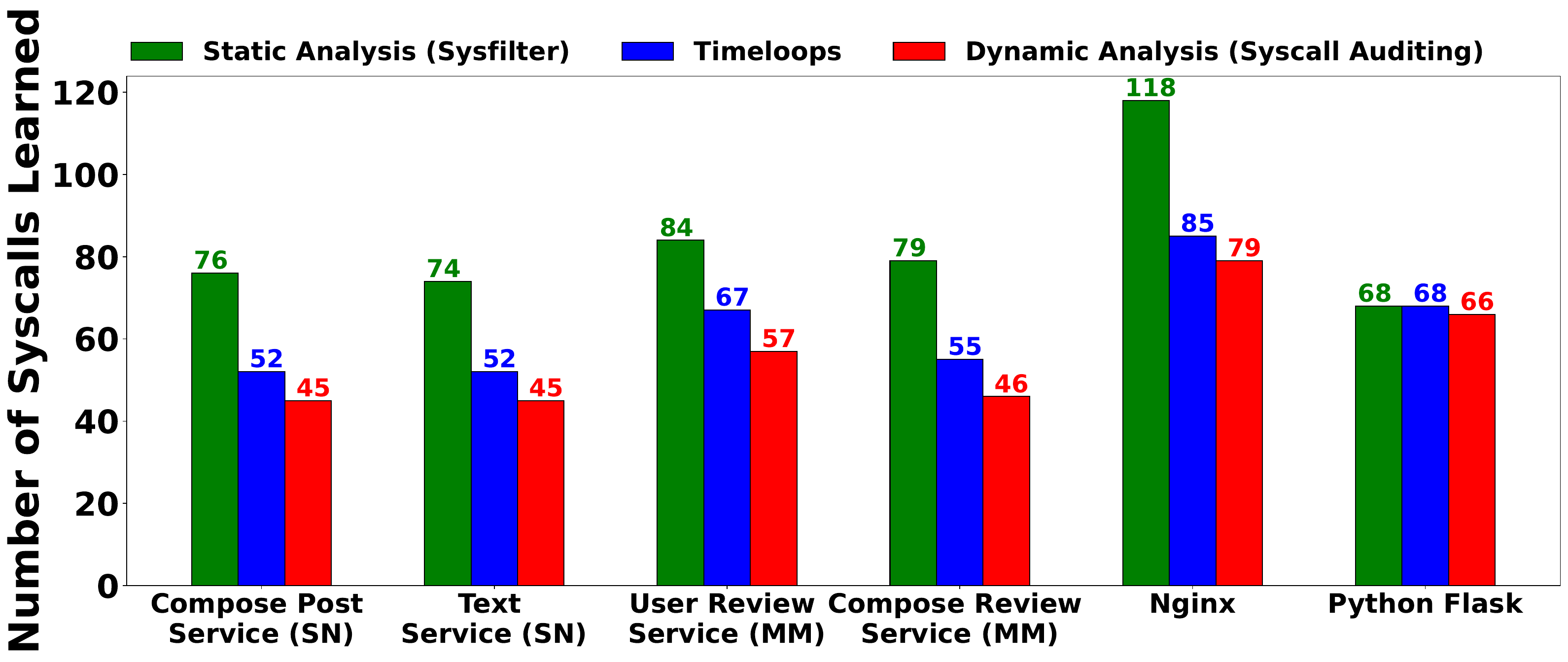}
    \caption{Comparison of the number of system calls allowed by static and
    dynamic system call filtering, compared to Timeloops.}
    \label{fig:syscalls-comparisons}
\end{figure}

In addition to the qualitative analysis, we evaluated the quality of the system
call policies created by \timeloops. To determine the effectiveness of
\timeloops for generating correct and safe policies, we sought to determine the
following:

\begin{enumerate}
    \item How many fewer system calls does \timeloops allow than
        over-approximate static analysis techniques (\textbf{RQ1})?
    \item Which system calls do static analysis techniques allow that are
        not allowed by \timeloops (\textbf{RQ2})?
    \item Which system calls does \timeloops allow that are not allowed by
        static analysis techniques (\textbf{RQ3})?
    \item Which system calls does \timeloops allow that would not have
        otherwise be necessary for the application to execute safely in a
        non-\timeloops system (\textbf{RQ4})?
\end{enumerate}

To evaluate the system call policies, we generated system call filters using
static analysis, \timeloops, and dynamic tracing. We chose \verb$sysfilter$
\cite{sysfilter} to represent state-of-the-art static analysis techniques for
system call policy creation. As a static analysis tool, \verb$sysfilter$ tends
to be over-approximate in its determinations. We used \verb$sysfilter$ to
generate syscall profiles for all our evaluated applications.

To generate our \timeloops system call policies, we executed each program it in
the \timeloops system with realistic workloads that exercised the functionality
of each service.

We additionally profiled the behavior of each evaluated program by executing it
in a container with \texttt{seccomp} \verb$RET_LOG$ mode, which allows, but
logs, each system call. We provided a set of sample inputs to the profiled
service and monitored the system calls invoked. This technique can be
considered elemental dynamic analysis for generating system call filters. This
provided a baseline of system calls that the program executes for a given input
without the introduction of new system calls from the \timeloops system.

\noindent~\textbf{RQ1\xspace}
For all evaluated programs, \verb$sysfilter$ created policies that were at
least as large as those created by \timeloops. On average, the \verb$sysfilter$
generated system call sets that were $32.7\%$ larger than those generated by
\timeloops, meaning that an attacker has access to approximately $32.7\%$ more
system calls when exploiting a system defended by \verb$sysfilter$ over one
defended by \timeloops, resulting in a larger attack surface and capability
set. Static approaches ignore the fact that inputs modulate program behavior,
at runtime, and estimate the system call behavior solely based on static
program code, thus over-approximating the set of allowed system calls. In
contrast, our results demonstrate that \timeloops is able to tailor the system
call list to specific inputs, and the degree of difference shows the typical
reduction in attack surface.

\noindent~\textbf{RQ2\xspace}
When comparing system call policies, we can quantify the size of the attack
surface with the number of system calls in a policy, but not all system calls
are created equally. Attackers do not need to access to \textit{every} system
call to be successful; they frequently require only a specific subset of system
calls that is unique to the post-exploitation behavior that they desire. This is
difficult to characterize, but in attempt to do so, we will analyze the policies
generated for two of the services that we evaluated: Nginx and the ComposePost
service from the Social Network application.

For both the Nginx and ComposePost services, \timeloops created smaller system
call policies than \verb$sysfilter$. The \timeloops filter for Nginx and
Compose Post were $40$ and $37$ system calls smaller than the \verb$sysfilter$
filter respectively. Many of these extraneous system calls were benign, like
\verb$exit$. Others, however, may provide opportunities to compromise the
container isolation from the host system, with numerous previous CVEs reported
with associated system calls in the Linux kernel. For example, \verb$sysfilter$
allowed both Nginx and ComposePost to have access to \verb$mremap$ (associated
with CVE-2020-10757), \verb$sendmmsg$ (CVE-2011-4594), and \verb$ftruncate$
(CVE-2018-18281), while \timeloops did not. \verb$sysfilter$ also allowed Nginx
to execute shared memory operations \verb$shmat$ and \verb$shmget$
(CVE-2017-5669), and allowed ComposePost to execute \verb$clock_settime$, while
\timeloops did not allow these. \verb$clock_settime$ not allowed by the default
Podman container filter due to its effects on the host system settings outside
of the container. Note that the presence of previous CVEs associated with a
system call is not enough to classify a system call as dangerous when executed
inside a container by an attacker; the CVEs simply show that there has been
some previous risk with allowing unrestricted access to the system call but do
not mean that the system call is still a risk, nor that a system call with no
associated CVE is \textit{not} a risk. We provide a full table of system calls
allowed by each system, along with a best-effort attempt to map the system
calls to associated Linux kernel CVEs, in Appendex \ref{sec:syscalltable}.

\noindent~\textbf{RQ3\xspace}
Some system calls are allowed by \timeloops that are not allowed by
\verb$sysfilter$, in part due to the ASan instrumentation applied to create the
oracle services. The Nginx \timeloops policy contained seven system calls that
were not present in the \verb$sysfilter$ policy, and the ComposePost \timeloops
policy contained $13$ system calls that were not present in the
\verb$sysfilter$. Some of these, like \verb$open$ (CVE-2020-8428
\cite{cve-2020-8428}) and \verb$pipe$ (CVE-2015-1805) do have history of being
abused in prior vulnerabilities, but overall this remains a smaller attacker
surface.

\noindent~\textbf{RQ4\xspace}
In order to determine which system calls were specifically introduced by the
\timeloops system, we compare the Nginx and ComposePost polices generated by
\timeloops to the system calls observed when executing the target program in a
container outside of the~\timeloops~system while profiling all system calls.
The \timeloops policies contain a superset of system calls compared to the
system calls observed in the baseline profile, meaning that all system calls
executed by the program outside of \timeloops were also in the \timeloops
policies. The \timeloops policies included six additional system calls in the
Nginx policy (\texttt{clock\_gettime}, \texttt{kill}, \texttt{madvise},
\texttt{open}, \texttt{readlink}, \texttt{sigaltstack}) and seven additional
system calls in the ComposePost policy (\texttt{getpid}, \texttt{gettid},
\texttt{readlink}, \texttt{sched\_getaffinity}, \texttt{sched\_yield},
\texttt{setrlimit}, \texttt{sigaltstack}). These system calls were likely
introduced by either ASan, as our oracle hardening technique, or by our
modifications to ensure that forking services properly exit when anomalous
activity is detected.

We conclude that~\timeloops~is effective at generating tight system call
policies when compared to current state-of-the-art static analysis approaches.
\timeloops does introduce some new system calls, but at a low incidence rate.

\subsection{Real-world Exploits}
\label{sec:sec-case-study}

We evaluate the effectiveness of \timeloops in the detection and prevention of
real-world exploits. Long-running web services are often subject to memory
errors that enable attackers to gain remote code execution abilities
\cite{malkiewicz_2021}. Buffer overflows, for example, account for many
software vulnerabilities and memory unsafe code remains the foundation for many
web services. We created a web server (written in C) to illustrate \timeloops's
ability to detect and prevent attacks that perform memory corruption and
strive to execute arbitrary code.

The web service contains a stack-based buffer overflow vulnerability modeled
after CVEs sampled throughout the last decade \cite{cve-2022-22274}
\cite{cve-2021-46393} \cite{cve-2009-0187} \cite{cve-2014-2206}. The vulnerable
web service serves a simple HTML page in response to an HTTP GET request.
Attackers can determine the addresses necessary to successfully exploit the
service and put together a payload, using a state-of-the-practice or
state-of-the-art code-reuse technique~\cite{shacham2007geometry,
checkoway2010return, bletsch2011jump, goktas2014out, bittau2014hacking,
schuster2015counterfeit, bosman2014framing, rudd2017address, snow2013just,
gawlik2016enabling, goktas2018position, goktas2020speculative,
mambretti2021bypassing}, which executes a TCP reverse shell. When the exploit
is run on an unprotected service, an attacker is able to successfully coerce
the service to ``connect back'' with a remote shell. However, when the attack
is executed in a system defended by \timeloops, the vulnerable buffer is
overwritten, but the attacker is unable to cause the service to connect back
(with a remote shell), because when the new system calls (\eg~\texttt{dup2},
\texttt{connect}) are executed to achieve this we switch to our oracle service
and detect the corresponding memory corruption. While this simple example can
be defended or patched in other ways, \timeloops's ability to mitigate
attacker's damage to the host machine can be extended to defending against the
entire class of remote code execution attacks.

Additionally, we evaluated the effectiveness of \timeloops on an Nginx
vulnerability CVE-2013-2028 \cite{cve-2013-2028}. This vulnerability allows an
attacker to abuse and integer signedness error and stack-based buffer to execute
arbitrary code. Similar to our exploit with the HTTP server, the exploit
succeeded on an undefended system, but was caught and prevented when \timeloops
system call filtering was enabled.

\subsection{Violating Retry-request Expectations}

The analysis conducted in Section~\ref{sec:attack-categories} expected that
identical requests are repeated until a valid response is sent from the
application. However, we must also consider the case where an attacker violates
this expectation or sends a different, malicious request instead of resending
the original input. If this happens, then the oracle and the production
services process different inputs, but this does not introduce any
opportunities to violate security properties of the system.

The oracle is tasked with determining whether a given input is safe or not.
Even if it is making the classification on an input different from the one that
the production service handled, it will still be able to make the determination
and update the security policy accordingly. If we supply a malicious request to
the production and a benign one to the oracle, the oracle will not add the new
malicious system calls to the filter and the malicious request will remain
incomplete. If we supply a benign request to the production that triggers the
oracle and a malicious one to the oracle, then the oracle will detect the
exploit and quit execution. If we supply two different malicious inputs to the
production and oracle services, the oracle will still be able to detect the
exploit from the second malicious request and the first one will remain
incomplete. There is no scenario where we can fool the oracle into adding a
system call to the allow-list with an exploit that the oracle is capable of
detecting.

This scenario still allows an attacker to abuse the system architecture to
cause targeted increased runtime overheads. Attackers can induce repeated
switches from application container to oracle by providing an input that
results in a new system call, but by then providing inputs that do not cause
system call violations, the oracle container handles all new requests but with
the higher performance overhead that comes from the exploit detection
instrumentation in the oracle. To prevent this from occurring (even
accidentally), we add a watchdog timer to the oracle container that causes it
to exit after some configurable amount of time, causing a switch back to the
more performant application container. Besides, this behavior is trivially
detectable using classic anomaly detection and/or prevention
techniques~\cite{chandola2009anomaly}.

\subsection{Application Container Compromise}

If an attacker achieves remote code execution in the application container, the
\timeloops~system is engineered to prevent full-system compromise.

First, \timeloops~leverages container abstractions provided by Podman to
isolate the application and oracle container resources. The containers are
restricted from accessing the full set of system calls by the container
\verb$seccomp-BPF$ policy that is set by \timeloops. The file representing the
system call policy is maintained entirely outside of the executing container
environments, and is only passed as an argument to Podman at container startup.
Therefore, an attacker with code execution inside either container is unable to
directly manipulate the policy file. Additionally, since \verb$seccomp-BPF$ is
handled in kernel space, all system call violations result in delivering
\texttt{SIGSYS} signals to the offending container's thread and recording the
violation in the system's audit log. It is possible for an attacker to
manipulate the userspace handling of the \texttt{SIGSYS} signals, given code
execution in the context of an application process. While this would prevent
\timeloops~from functioning correctly, it would not allow the attacker to
loosen the existing syscall policy which is immutable (kernel-enforced).

We further mitigate the danger of a successful remote attacker by running all
containers without administrative privileges. This ensures that any container
escapes result in an attacker having the same permissions as the
\timeloops~user, and no more. These limited permissions do not allow attackers
to dismantle the system call filter or alter running processes on the host. We
believe that due to the series of security benefits provided by our approach,
from exploit detection and tight system call filtering to container isolation,
\timeloops sufficiently raises the bar for successful system compromise.

\subsection{\rtimeloops vs. Permanent Hardening}

We compare the security guarantees that \timeloops~provides with a permanently
hardened system, or in our case, a service that is always running with ASan. In
the \timeloops~design, we only check inputs with ASan when they trigger new
system calls. In the case that a malicious input does not invoke a new system
call, but induces memory corruption, our \timeloops~production service will be
corrupted, but an ASan-hardened service will catch the exploit and quit.  We
only offer protection equal to an ASan-hardened service when new system calls
are invoked, but we make the key observation that malicious payloads tend to
invoke new system calls. By design, our hardened oracle service will inspect
those inputs with greater scrutiny. Since it is impractical to deploy
ASan-hardened services due to their great performance overheads, our approach
provides the security benefits of ASan-hardening without incurring the costs.
\timeloops enables heavyweight security techniques to be practical in
performance-critical deployments.

\section{Performance Evaluation}
\label{sec:performance}

To evaluate the performance of \timeloops, we consider the end-to-end latency
of network requests sent to a service application that is running under our
\timeloops prototype. We measure latency as the elapsed time between a client
making the first request to the service and the client receiving a response,
regardless of the number of times the client may have retried the request.

We compare the latency of the application running under \timeloops to the
latency of the same application running with no additional exploit mitigations
applied (\ie~outside of \timeloops). We additionally compare the latency of the
\timeloops-hardened application to that of the same application executing with
ASan-hardening, outside of \timeloops. This represents a system that is
instrumented to maximally determine when an input causes a security violation.
Our evaluation aims to show that \timeloops~can monitor for violations of
security boundaries while also providing amortized performance benefits that
are comparable to that of an unhardened (and hence unprotected) system.

\begin{figure*}
    \centering

    \begin{subfigure}[t]{0.35\textwidth}
        \centering
        \includegraphics[width=\linewidth]{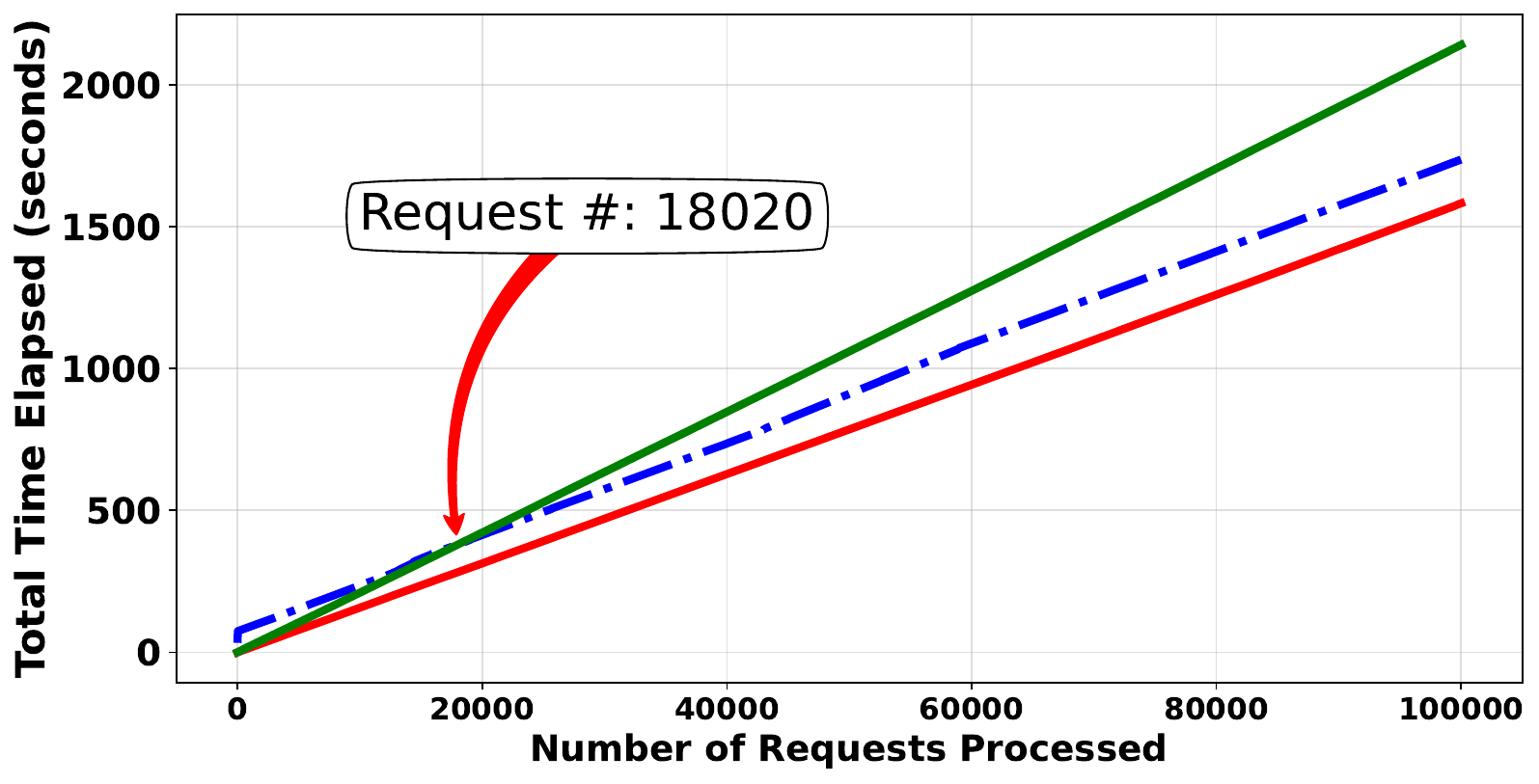}
        \caption{Compose Review (MM)}
        \label{fig:compose-review-mm-latency}
    \end{subfigure}  \qquad
    \begin{subfigure}[t]{0.35\textwidth}
        \centering
        \includegraphics[width=\linewidth]{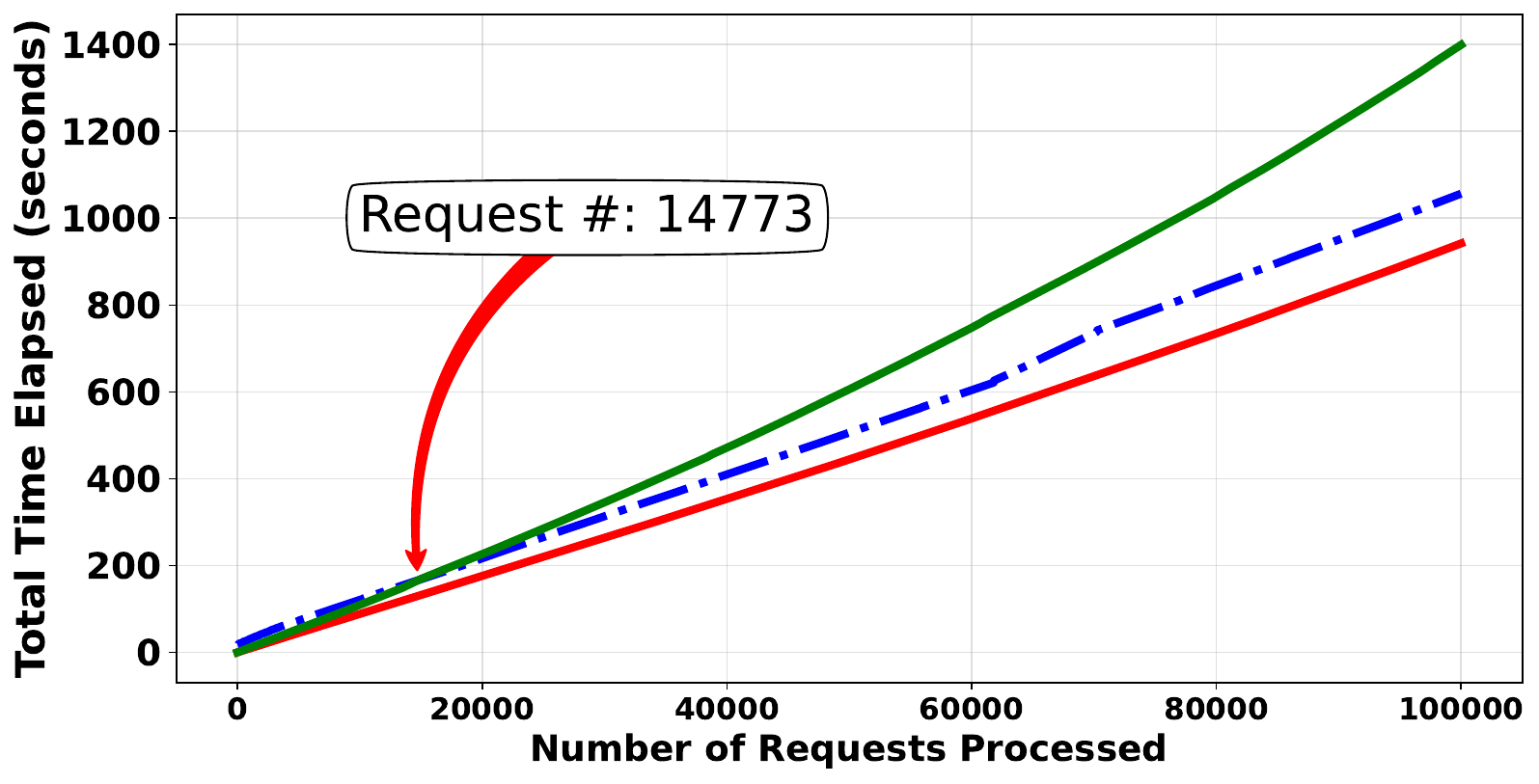}
        \caption{Compose Post (SN)}
        \label{fig:compose-post-sn-latency}
    \end{subfigure} \qquad
    \begin{subfigure}[t]{0.35\textwidth}
        \centering
        \includegraphics[width=\linewidth]{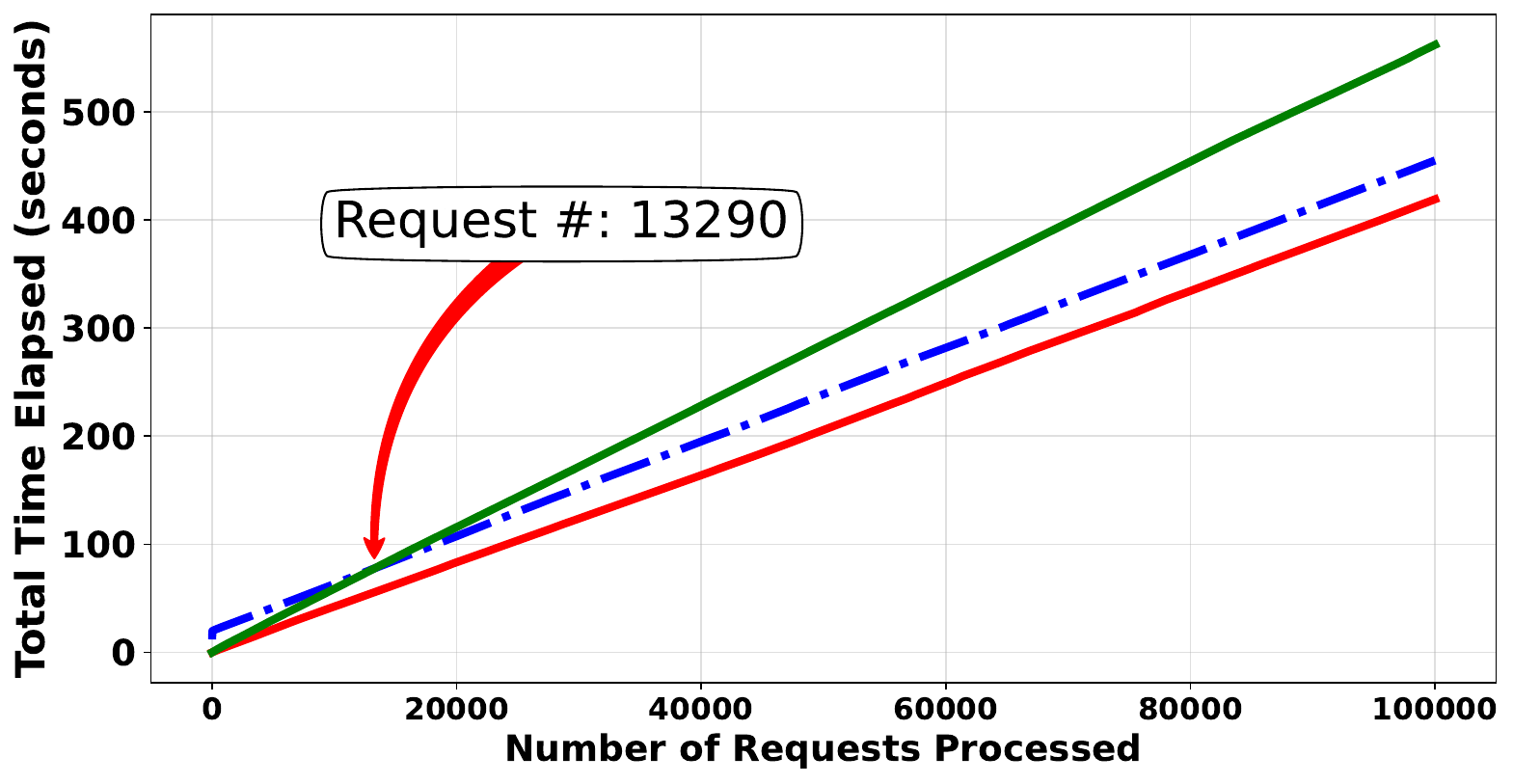}
        \caption{Python Flask}
        \label{fig:flask-performance-latency}
    \end{subfigure}  \qquad
    \begin{subfigure}[t]{0.35\textwidth}
        \centering
        \includegraphics[width=\linewidth]{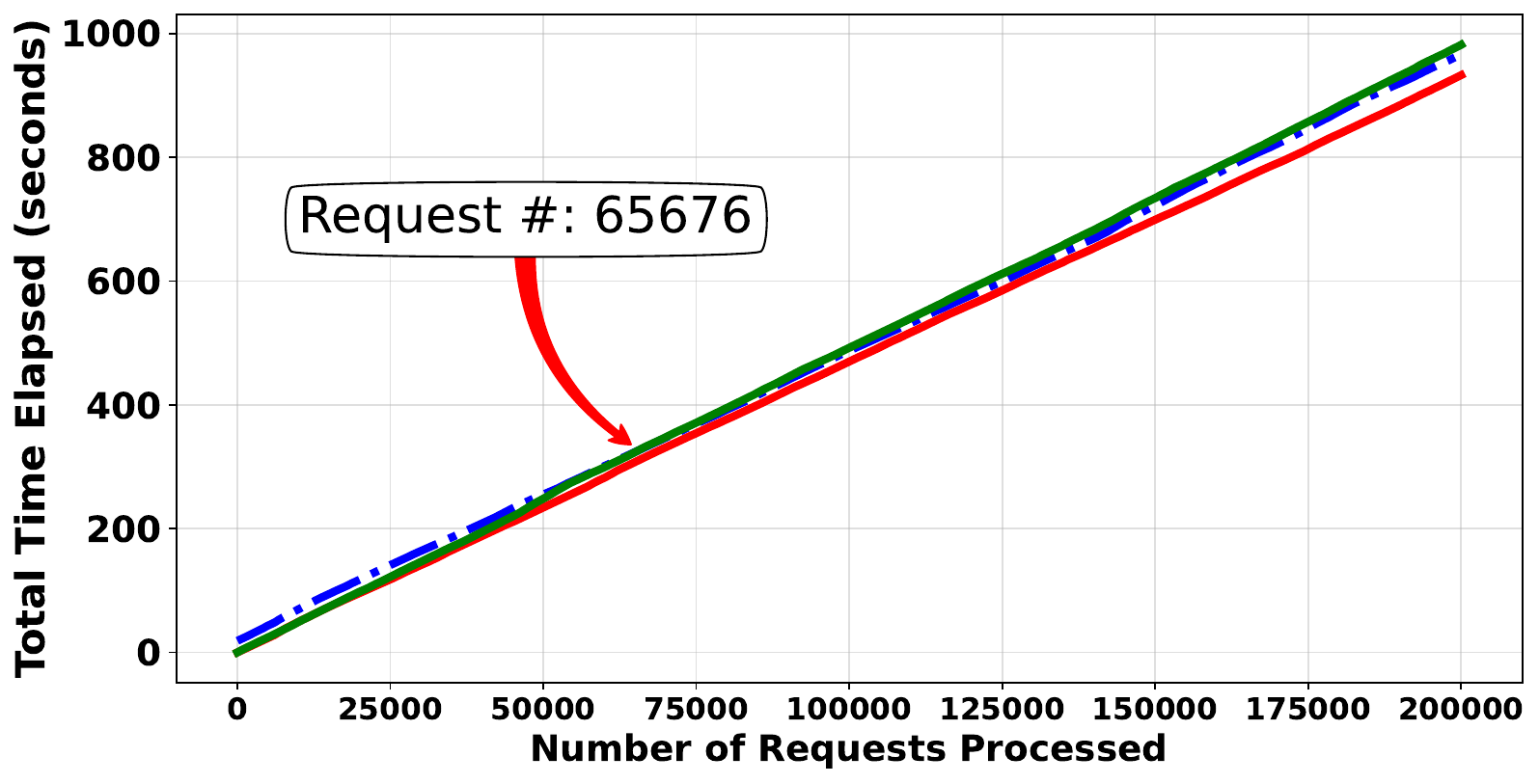}
        \caption{Nginx}
        \label{fig:nginx-performance-latency}
    \end{subfigure}  \qquad
    \begin{subfigure}[t]{0.4\textwidth}
        \centering
        \includegraphics[width=\linewidth]{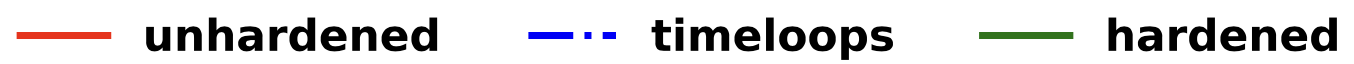}
        \label{fig:legend}
    \end{subfigure}  \qquad
       \caption{Comparison of cumulative time for processing repeated requests
           sent to an unhardened service, a \timeloops service, and a hardened
           service. Figures \ref{fig:compose-review-mm-latency} and
           \ref{fig:compose-post-sn-latency} show the end-to-end latency of
           sending requests to the media microservices (MM) and social network
           microservices (SN) benchmark. Figures
           \ref{fig:flask-performance-latency} and
           \ref{fig:nginx-performance-latency} show the end-to-end latency of
           sending requests to each of the stand-alone applications that process
           web requests. In every case, the \timeloops service starts with a
           longer response time while system calls are learned, but eventually
           overtakes the hardened service (intercept is plotted).}
       \label{fig:latency-timeseries}
\end{figure*}

\begin{figure*}
    \centering

    \begin{subfigure}[t]{0.35\textwidth}
        \centering
        \includegraphics[width=\textwidth]{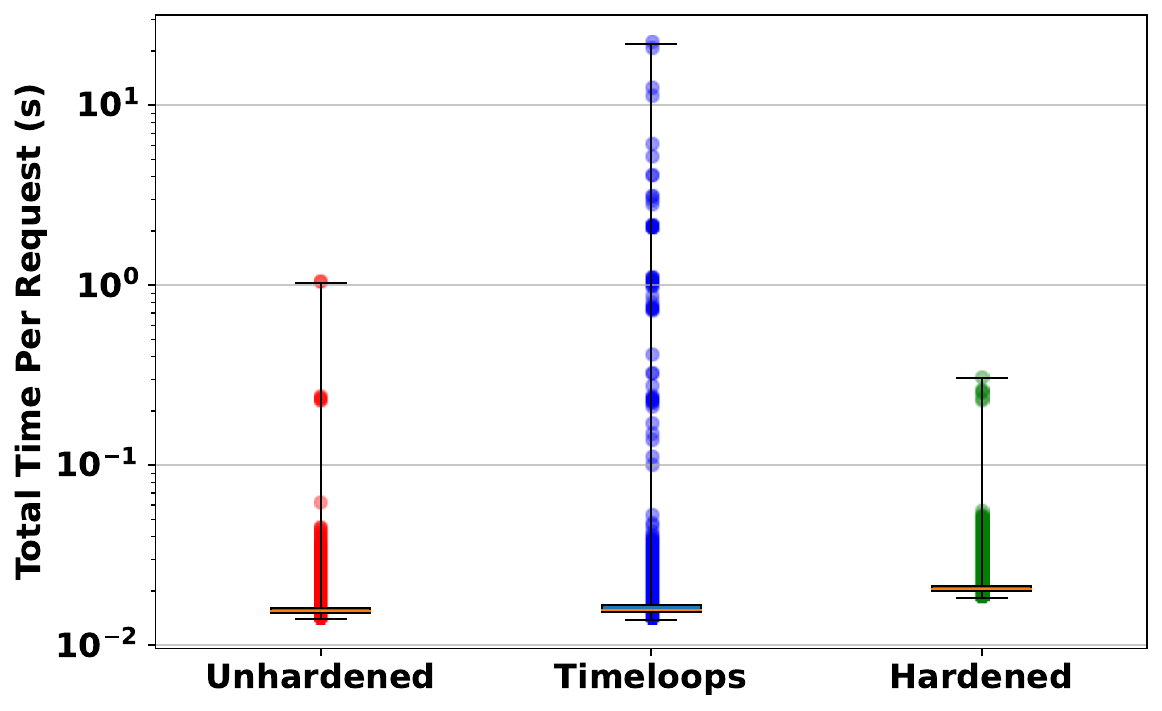}
        \caption{Compose Review (MM)}
        \label{fig:compose-review-boxplot}
    \end{subfigure}  \qquad
    \begin{subfigure}[t]{0.35\textwidth}
        \centering
        \includegraphics[width=\textwidth]{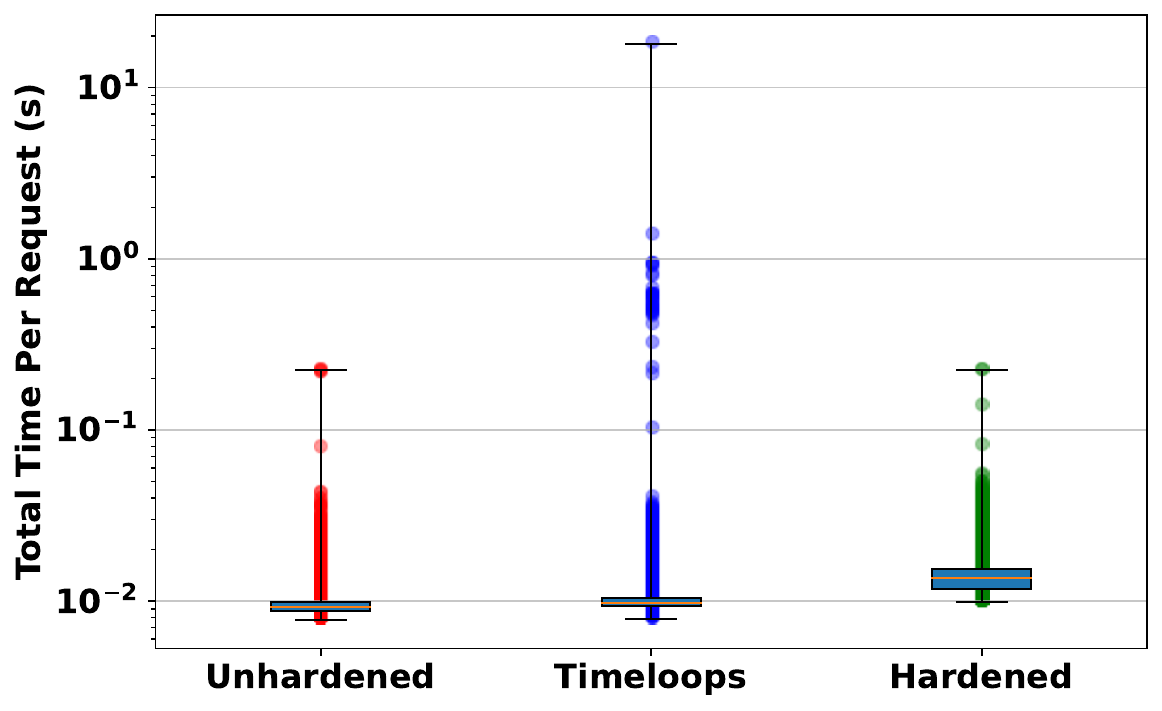}
        \caption{Compose Post (SN)}
        \label{fig:compose-post-sn-boxplot}
    \end{subfigure} \qquad
    \begin{subfigure}[t]{0.35\textwidth}
        \centering
        \includegraphics[width=\textwidth]{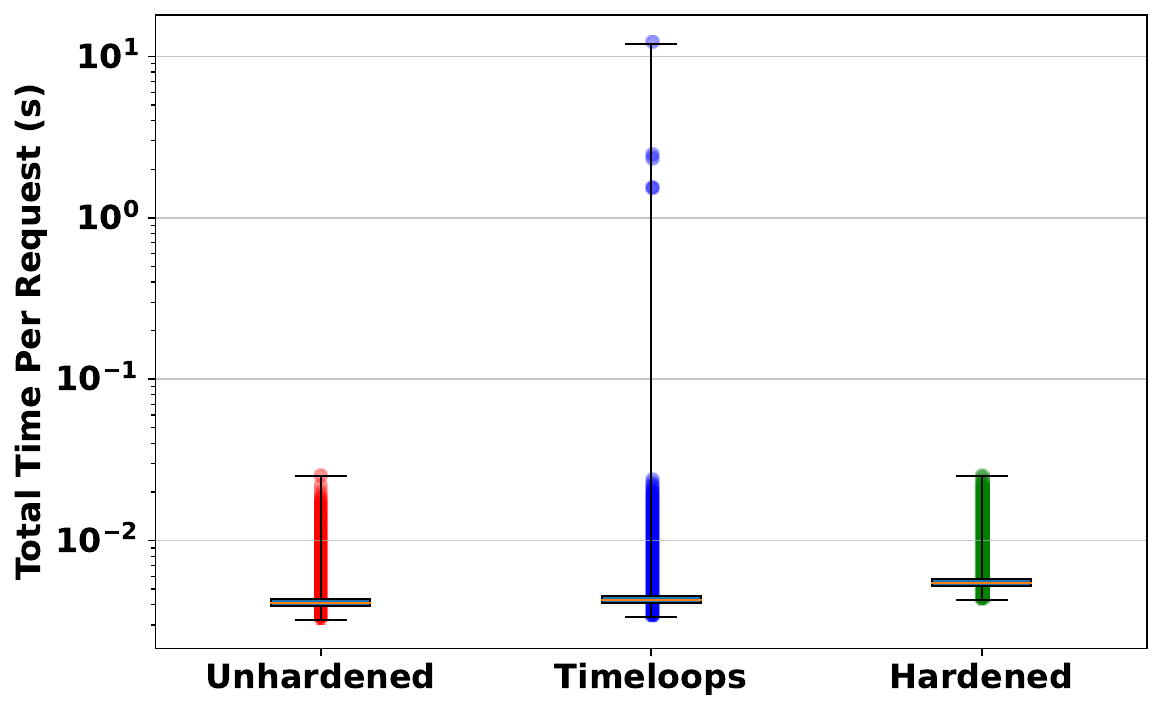}
        \caption{Python Flask}
        \label{fig:python-boxplot}
    \end{subfigure}  \qquad
    \begin{subfigure}[t]{0.35\textwidth}
        \centering
        \includegraphics[width=\textwidth]{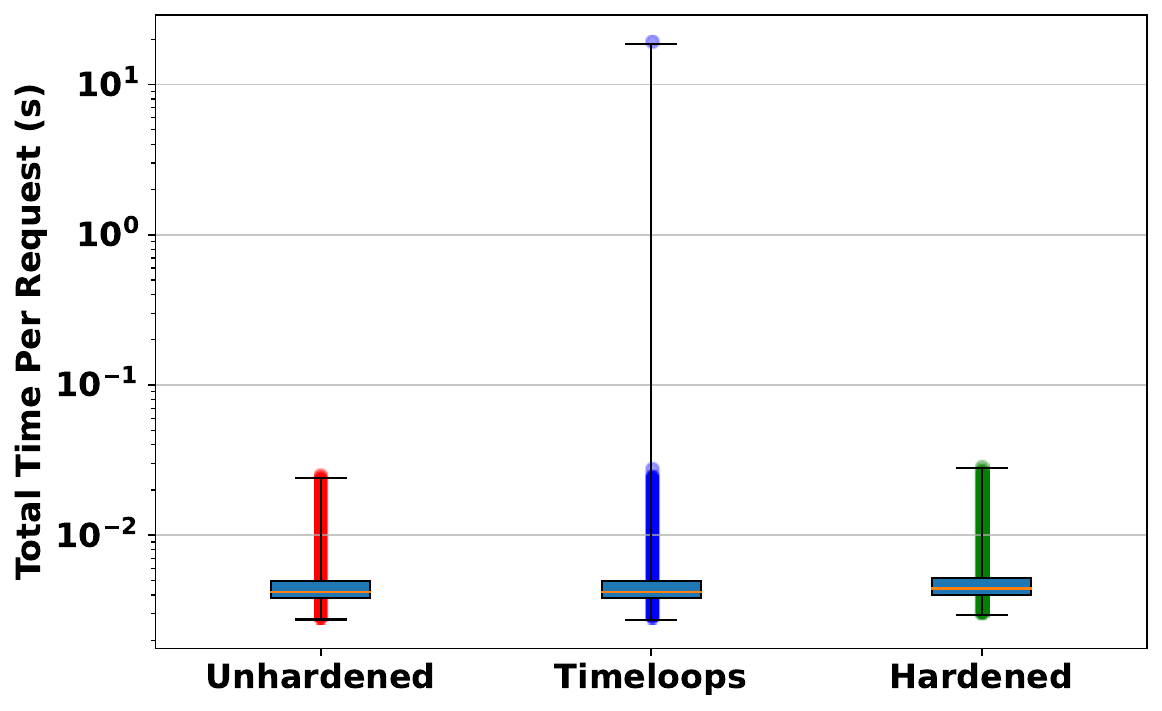}
        \caption{Nginx}
        \label{fig:nginx-boxplot}
    \end{subfigure}  \qquad
       \caption{Comparison of request latency distribution per request sent to
           an unhardened service, a \timeloops service, and a hardened service.
           The average request latencies of the \timeloops services are similar
           to that of the unhardened service and lower than the hardened
           service, despite the maximum value of the \timeloops service being a
           significant outlier. The maximum value comes from the large delay in
           servicing the first requests as system calls are learned.}
       \label{fig:latency-boxplots}
\end{figure*}

All experiments began by executing the test application service in \timeloops
with an empty system call allow-list. We used an HTTP client program to send
each request to the application repeatedly until it was successfully processed.
Some system calls are iteratively learned even before the application is able
to receive its first input. 

In all experiments, the very first request sent to the \timeloops application
resulted in very slow response times since it typically required learning many
new system calls. Following requests tended not to invoke many new system calls
and were therefore quickly processed. Figure~\ref{fig:latency-timeseries} shows
the latency of each request sent, and how the application running in
\timeloops~always has a slower start than the other services. However, the
average latency, shown in Figure~\ref{fig:latency-boxplots}, of ensuing requests
processed by the \timeloops~application was close to that of the completely
unhardened service. In all trials the average \timeloops~request outperformed
the average hardened service request.

The high latency of the first few requests led to another observation: by
pre-training \timeloops~on known-safe inputs to the application, we can create
an initial allow-list prior to deployment. However, unlike in traditional
dynamic system call learning systems, these initial inputs do not need to
exhaustively include every system call that should be learned---in this case,
the \timeloops~system can still learn new system calls from future inputs. By
training on the most performance-critical inputs ahead of time, an operator of
the \timeloops~system can further amortize system call learning costs on the
critical path.

\section{Related Work}
\label{sec:relatedwork}

\subsection{Static System Call Filtering}

Static system call filtering approaches attempt to determine a correct system
call policy without executing the program, and then apply the policy via some
enforcement mechanism. \verb$sysfilter$ \cite{sysfilter} is a binary
analysis-based framework that synthesizes syscall filters and applies them to
applications. Like our approach, \verb$sysfilter$ determines developer-intended
program behavior and enforces only that behavior via \verb$seccomp-BPF$
filters. However, \verb$sysfilter$ determines this behavior statically and
analyzes call graphs to generate their system call filters. While the authors
use many techniques to prune this call graph to avoid adding unreachable
syscalls to their filter, it still may be too large. Abhaya \cite{abhaya}, like
\verb$sysfilter$, uses static code analysis to generate system call policies,
but analyzes program source code, rather than compiled binaries, and targets
both \verb$seccomp-BPF$ and Pledge policies for enforcement.

Temporal syscall filtering \cite{temp-sys-filtering} relies on static analysis
to create syscall filters for applications. It differs from \verb$sysfilter$
because it makes the observation that many system calls that are used for the
initialization of an application are no longer required during the
application’s steady state. The authors create two filters: one that includes
all of the initialization syscalls, applied during the ``init'' execution
phase, while the other is applied during the ``main loop'' phase.

Saphire \cite{bulekov2021saphire} presents an approach for creating system call
policies for interpreted languages. Specifically, Saphire creates sandboxes for
PHP applications by analyzing PHP source code rather than attempting to analyze
the entire interpreter.

\subsection{Dynamic System Call Filtering}

Dynamic syscall filtering tools execute the target program various inputs to
observe which system calls are required by an application. Like \timeloops,
dynamic approaches execute the program to derive policies based on observations
of program behavior. However, they tend to execute in distinct phases of
training and deployment, while \timeloops conducts its learning while able to
service production workloads. ZenIDS \cite{zenids} is able to learn system call
policies for PHP applications with an online training period to then monitor
for anomalies. Systrace \cite{systrace} also uses dynamic tracing to learn
policies, but enforces the policies by implementing a userspace daemon.

\subsection{Cloud and Container Security}

As cloud and container environments have proliferated, new approaches have been
proposed to secure them. Confine \cite{ghavamnia2020confine} is a static system
call filtering technique specifically for analyzing the contents of containers
to determine allowed system calls exposed to the container. Dynamic approaches
have also been applied to containers \cite{mining-sandboxes} to profile the
syscalls invoked by the container during the execution of automated test
inputs, and to enforce a policy restricted by the profile. Work has been done
to analyze and categorize \cite{linux-container-security} privilege escalation
container exploits that break the resource isolation provided by container
abstractions.  AUTOARMOR \cite{li2021automatic} is a proposal to automatically
generate policies for securing microservices by analyzing the interactions
between them.

\section{Future Work}\label{sec:futurework}

\subsection{Optimizations}\label{sec:optimizations}

\subsubsection{Performance Optimizations}
In our evaluation, we saw that most of the additional overheads that timeloops
applications have in comparison to the unhardened services come from processing
the first few requests. The rest of the execution after this "warmup phase" is
similar to that of the unhardened services. To avoid incurring this warmup cost,
we can save filters and reuse them in the future if we are confident of thier 
correctness and can validate that they have not been compromised between uses. 

Additionally, we see that it is quite costly to stop and restart containers, but
this step is necessary since Seccomp filters are immutable once installed. This
is a feauture of Seccomp since we do not want attackers to be able to modify the
Seccomp filters that are installed on a running program. This could lead to
uninstalling the Seccomp filter entirely or adding system calls necessary for an
attack. If we had a custom eBPF program that allowed our trusted timeloops
infrastructure to modify the system call filter, then we could avoid incurring
this cost every time we encounter a new system call. 

\subsubsection{Security Optimizations}
Our current implementation of \timeloops involves iteratively writing a security
policy by creating an allow-list of behaviors. However, over time, the intended
behavior of an application may change depending on the phase of execution, the
set of inputs, the time of day or some other factor. When application behavior
changes, we do not want to include old behaviors in our security policy any
longer. To further tighten our security guarantees, we could implement periodic
policy forgetting. This would allow us to ensure that our policy enforces tight
bounds and only encapsulates current behavior. 

Another assumption we make in our timeloops design is that malicious behavior is
always introduced via a vulnerability that our oracle can detect. This is not
always the case. Future instantiations of \timeloops could use different or even
multiple hardening techniques for the oracle. 

Another optimization for our system call learning case study is to keep track of
system call arguments. This will enforce a finer-grained security policy. Even
though common system calls are used during attacks, they are often invoked with
different arguments than the ones used in normal program execution.

\begin{figure}[t]
    \includegraphics[width=\linewidth]{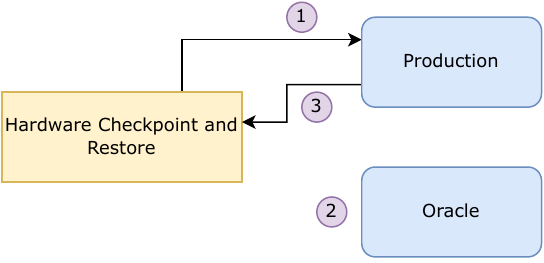}
    \caption{With additional hardware checkpointing and replaying support, when a new syscall is encountered we will (1) save the state of the production service and pause execution, (2) execute the request in the oracle and update the policy, and (3) restore the production service state from the checkpoint and resume execution.}
    \label{fig:syscalls-comparisons}
\end{figure}

\subsubsection{Hardware Support}
Introducing hardware support in our \timeloops design may allow us to
incorporate some of these optimizations with ease. For example, hardware
checkpointing techniques \cite{efficient-hw} offer the ability to simply pause
execution in one instance of a service and resume execution later. Instead of
resending a request to the production service again, we can simply resume
execution. Hardware checkpointing in conjunction with hardware support for
replaying execution could save us the overheads of stopping services each time
we need to consult the oracle. This support could potentially allow us to
implement timeloops on services that are not stateless as well. 
\section{Conclusion}
\label{sec:conclusion}

Creating security policies for microservices is a difficult and tedious
process: it requires a thorough understanding of each service and must flexibly
adapt to changes in the deployment. This paper takes the first steps towards
automatically crafting security policies for containerized microservices. We
aim to learn policies for system call filtering, a technology that is becoming
increasingly necessary for maintaining isolation in cloud settings.

To learn policies automatically, we introduced the concept of timelooping. The
key idea is to: (i)~re-execute a program among two variants of an application,
one hardened for security, and another one optimized for performance;
(ii)~learn execution properties from the hardened version; and (iii)~use the
results of the previous step to craft policies that can be enforced on the
performance-optimized version. Our solution takes a pragmatic stance in the
trade-off(s) between security and performance.

We demonstrated the merit and applicability of our \timeloops learning scheme
in three significant ways: (1)~we showed that the system call allow-lists
created by \timeloops for our applications are significantly better than
statically- and dynamically-generated policies, (2)~we showed that the idea is
compatible-with, and well-suited to, modern software engineering practices,
such the use of containerized microservices crafted using popular interpreted
languages. These use cases have been the Achilles' heel of several security
studies in our community, and (3)~we showed that the idea can be easily
implemented by combining existing technologies.

In our community there has been significant amount of work on hardening
applications. As is to be expected, these hardening services come with
performance and energy overheads. In an ideal world, these overheads should not
matter. In reality, however, users care about performance, forcing them to make
a hard choice between performance and security. In our \timeloops system we use
the oracle version as needed to learn security properties, and then enforce
these security properties on the production version using lightweight
enforcement techniques. Thus, \timeloops obviates the need to make this
performance-security choice. With time, we expect properties and policies
beyond system call filtering to use our \timeloops technique.

\section{Acknowledgements and Disclosures}\label{sec:acknowledgements}

This work was partially supported by N00014-20-1-2746, a NSF Graduate Research
Fellowship, and a gift from Bloomberg. Any opinions, findings, conclusions and
recommendations expressed in this material are those of the authors and do not
necessarily reflect the views of the US government or commercial entities. Simha
Sethumadhavan has a significant financial interest in Chip Scan Inc. Patent
Pending.

\bibliographystyle{plain}
\bibliography{main}

\begin{thebibliography}{10}

\bibitem{cve-2009-0187}
Cve-2009-0187.
\newblock National Vulnerability Database.

\bibitem{cve-2013-2028}
Cve-2013-2028.
\newblock National Vulnerability Database.

\bibitem{cve-2014-2206}
Cve-2014-2206.
\newblock National Vulnerability Database.

\bibitem{cve-2020-8428}
Cve-2020-8428.
\newblock National Vulnerability Database.

\bibitem{cve-2021-46393}
Cve-2021-46393.
\newblock National Vulnerability Database.

\bibitem{cve-2022-22274}
Cve-2022-22274.
\newblock National Vulnerability Database.

\bibitem{podman}
Pod manager tool (podman).
\newblock podman website.

\bibitem{ec2}
What is amazon ec2?
\newblock AWS User Guide for Linux Instances.

\bibitem{host-security}
Improving host security with system call policies.
\newblock In {\em 12th USENIX Security Symposium (USENIX Security 03)},
  Washington, D.C., August 2003. USENIX Association.

\bibitem{dockerdocs}
Seccomp security profiles for docker, Dec 2021.

\bibitem{abubakar2021shard}
Muhammad Abubakar, Adil Ahmad, Pedro Fonseca, and Dongyan Xu.
\newblock $\{$SHARD$\}$: Fine-grained kernel specialization with context-aware
  hardening.
\newblock In {\em 30th $\{$USENIX$\}$ Security Symposium ($\{$USENIX$\}$
  Security 21)}, 2021.

\bibitem{bittau2014hacking}
Andrea Bittau, Adam Belay, Ali Mashtizadeh, David Mazi{\`e}res, and Dan Boneh.
\newblock {Hacking Blind}.
\newblock In {\em IEEE Symposium on Security and Privacy (S\&P)}, pages
  227--242, 2014.

\bibitem{bletsch2011jump}
Tyler Bletsch, Xuxian Jiang, Vince~W Freeh, and Zhenkai Liang.
\newblock {Jump-Oriented Programming: A New Class of Code-Reuse Attack}.
\newblock In {\em ACM Asia Symposium on Information, Computer and
  Communications Security (ASIACCS)}, pages 30--40, 2011.

\bibitem{bosman2014framing}
Erik Bosman and Herbert Bos.
\newblock {Framing Signals---A Return to Portable Shellcode}.
\newblock In {\em IEEE Symposium on Security and Privacy (S\&P)}, pages
  243--258, 2014.

\bibitem{bulekov2021saphire}
Alexander Bulekov, Rasoul Jahanshahi, and Manuel Egele.
\newblock Saphire: Sandboxing $\{$PHP$\}$ applications with tailored system
  call allowlists.
\newblock In {\em 30th $\{$USENIX$\}$ Security Symposium ($\{$USENIX$\}$
  Security 21)}, 2021.

\bibitem{burow2019sok}
Nathan Burow, Xinping Zhang, and Mathias Payer.
\newblock Sok: Shining light on shadow stacks.
\newblock In {\em 2019 IEEE Symposium on Security and Privacy (SP)}, pages
  985--999. IEEE, 2019.

\bibitem{canella2021automating}
Claudio Canella, Mario Werner, Daniel Gruss, and Michael Schwarz.
\newblock Automating seccomp filter generation for linux applications.
\newblock In {\em Proceedings of the 2021 on Cloud Computing Security
  Workshop}, pages 139--151, 2021.

\bibitem{chestnut}
Claudio Canella, Mario Werner, Daniel Gruss, and Michael Schwarz.
\newblock Automating seccomp filter generation for linux applications.
\newblock CCSW '21, page 139–151, New York, NY, USA, 2021. Association for
  Computing Machinery.

\bibitem{containerescapology}
Canister.
\newblock Container escapes: An exercise in practical container escapology, Mar
  2019.

\bibitem{chandola2009anomaly}
Varun Chandola, Arindam Banerjee, and Vipin Kumar.
\newblock Anomaly detection: A survey.
\newblock {\em ACM computing surveys (CSUR)}, 41(3):1--58, 2009.

\bibitem{checkoway2010return}
Stephen Checkoway, Lucas Davi, Alexandra Dmitrienko, Ahmad-Reza Sadeghi, Hovav
  Shacham, and Marcel Winandy.
\newblock {Return-Oriented Programming without Returns}.
\newblock In {\em ACM Conference on Computer and Communications Security
  (CCS)}, pages 559--572, 2010.

\bibitem{chierici2022}
Stefano Chierici.
\newblock Cve-2022-0492: Privilege escalation vulnerability causing container
  escape, Mar 2022.

\bibitem{cockcroftms}
Adrian Cockcroft.
\newblock Evolution of microservices - craft conference, Apr 2016.

\bibitem{tail-at-scale}
Jeffrey Dean and Luiz~Andr\'{e} Barroso.
\newblock The tail at scale.
\newblock {\em Commun. ACM}, 56(2):74–80, feb 2013.

\bibitem{sysfilter}
Nicholas DeMarinis, Kent Williams-King, Di~Jin, Rodrigo Fonseca, and
  Vasileios~P. Kemerlis.
\newblock sysfilter: Automated system call filtering for commodity software.
\newblock In {\em 23rd International Symposium on Research in Attacks,
  Intrusions and Defenses ({RAID} 2020)}, pages 459--474, San Sebastian,
  October 2020. {USENIX} Association.

\bibitem{dockerseccomp}
{Docker Documentation}.
\newblock Seccomp security profiles for docker.

\bibitem{edprice-msft}
EdPrice-MSFT.
\newblock N-tier architecture style - azure architecture center.

\bibitem{dsb}
Yu~Gan, Yanqi Zhang, Dailun Cheng, Ankitha Shetty, Priyal Rathi, Nayan Katarki,
  Ariana Bruno, Justin Hu, Brian Ritchken, Brendon Jackson, Kelvin Hu, Meghna
  Pancholi, Yuan He, Brett Clancy, Chris Colen, Fukang Wen, Catherine Leung,
  Siyuan Wang, Leon Zaruvinsky, Mateo Espinosa, Rick Lin, Zhongling Liu, Jake
  Padilla, and Christina Delimitrou.
\newblock An open-source benchmark suite for microservices and their
  hardware-software implications for cloud \&; edge systems.
\newblock In {\em Proceedings of the Twenty-Fourth International Conference on
  Architectural Support for Programming Languages and Operating Systems},
  ASPLOS '19, page 3–18, New York, NY, USA, 2019. Association for Computing
  Machinery.

\bibitem{gawlik2016enabling}
Robert Gawlik, Benjamin Kollenda, Philipp Koppe, Behrad Garmany, and Thorsten
  Holz.
\newblock {Enabling Client-Side Crash-Resistance to Overcome Diversification
  and Information Hiding}.
\newblock In {\em Network and Distributed System Security Symposium (NDSS)},
  2016.

\bibitem{ghavamnia2020confine}
Seyedhamed Ghavamnia, Tapti Palit, Azzedine Benameur, and Michalis
  Polychronakis.
\newblock Confine: Automated system call policy generation for container attack
  surface reduction.
\newblock In {\em 23rd International Symposium on Research in Attacks,
  Intrusions and Defenses ($\{$RAID$\}$ 2020)}, pages 443--458, 2020.

\bibitem{temp-sys-filtering}
Seyedhamed Ghavamnia, Tapti Palit, Shachee Mishra, and Michalis Polychronakis.
\newblock Temporal system call specialization for attack surface reduction.
\newblock In {\em 29th {USENIX} Security Symposium ({USENIX} Security 20)},
  pages 1749--1766. {USENIX} Association, August 2020.

\bibitem{goktas2014out}
Enes G{\"o}ktas, Elias Athanasopoulos, Herbert Bos, and Georgios Portokalidis.
\newblock {Out Of Control: Overcoming Control-Flow Integrity}.
\newblock In {\em IEEE Symposium on Security and Privacy (S\&P)}, pages
  575--589, 2014.

\bibitem{goktas2018position}
Enes G{\"o}ktas, Benjamin Kollenda, Philipp Koppe, Erik Bosman, Georgios
  Portokalidis, Thorsten Holz, Herbert Bos, and Cristiano Giuffrida.
\newblock {Position-independent Code Reuse: On the Effectiveness of ASLR in the
  Absence of Information Disclosure}.
\newblock In {\em IEEE European Symposium on Security and Privacy (EuroS\&P)},
  pages 227--242, 2018.

\bibitem{goktas2020speculative}
Enes G{\"o}ktas, Kaveh Razavi, Georgios Portokalidis, Herbert Bos, and
  Cristiano Giuffrida.
\newblock {Speculative Probing: Hacking Blind in the Spectre Era}.
\newblock In {\em ACM Conference on Computer and Communications Security
  (CCS)}, pages 1871--1885, 2020.

\bibitem{zenids}
Byron Hawkins and Brian Demsky.
\newblock Zenids: Introspective intrusion detection for php applications.
\newblock In {\em 2017 IEEE/ACM 39th International Conference on Software
  Engineering (ICSE)}, pages 232--243, 2017.

\bibitem{hedgeskeating2021}
Matt Hedges and Joseph Keating.
\newblock Deploying python flask microservices to aws using open source tools,
  Apr 2021.

\bibitem{kim2021prof}
Sungjin Kim, Byung~Joon Kim, and Dong~Hoon Lee.
\newblock Prof-gen: Practical study on system call whitelist generation for
  container attack surface reduction.
\newblock In {\em 2021 IEEE 14th International Conference on Cloud Computing
  (CLOUD)}, pages 278--287. IEEE, 2021.

\bibitem{efficient-hw}
Dirk Koch, Christian Haubelt, and J\"{u}rgen Teich.
\newblock Efficient hardware checkpointing: Concepts, overhead analysis, and
  implementation.
\newblock In {\em Proceedings of the 2007 ACM/SIGDA 15th International
  Symposium on Field Programmable Gate Arrays}, FPGA '07, page 188–196, New
  York, NY, USA, 2007. Association for Computing Machinery.

\bibitem{spectre}
Paul Kocher, Jann Horn, Anders Fogh, Daniel Genkin, Daniel Gruss, Werner Haas,
  Mike Hamburg, Moritz Lipp, Stefan Mangard, Thomas Prescher, et~al.
\newblock Spectre attacks: Exploiting speculative execution.
\newblock {\em Communications of the ACM}, 63(7):93--101, 2020.

\bibitem{larsen2014sok}
Per Larsen, Andrei Homescu, Stefan Brunthaler, and Michael Franz.
\newblock Sok: Automated software diversity.
\newblock In {\em 2014 IEEE Symposium on Security and Privacy}, pages 276--291.
  IEEE, 2014.

\bibitem{li2021automatic}
Xing Li, Yan Chen, Zhiqiang Lin, Xiao Wang, and Jim~Hao Chen.
\newblock Automatic policy generation for inter-service access control of
  microservices.
\newblock In {\em 30th $\{$USENIX$\}$ Security Symposium ($\{$USENIX$\}$
  Security 21)}, 2021.

\bibitem{li2017lock}
Yiwen Li, Brendan Dolan-Gavitt, Sam Weber, and Justin Cappos.
\newblock Lock-in-pop: Securing privileged operating system kernels by keeping
  on the beaten path.
\newblock In {\em 2017 $\{$USENIX$\}$ Annual Technical Conference
  ($\{$USENIX$\}$$\{$ATC$\}$ 17)}, pages 1--13, 2017.

\bibitem{linux-container-security}
Xin Lin, Lingguang Lei, Yuewu Wang, Jiwu Jing, Kun Sun, and Quan Zhou.
\newblock A measurement study on linux container security: Attacks and
  countermeasures.
\newblock In {\em Proceedings of the 34th Annual Computer Security Applications
  Conference}, ACSAC '18, page 418–429, New York, NY, USA, 2018. Association
  for Computing Machinery.

\bibitem{manseccomp}
{Linux Programmer's Manual}.
\newblock Seccomp(2).

\bibitem{meltdown}
Moritz Lipp, Michael Schwarz, Daniel Gruss, Thomas Prescher, Werner Haas, Jann
  Horn, Stefan Mangard, Paul Kocher, Daniel Genkin, Yuval Yarom, et~al.
\newblock Meltdown: Reading kernel memory from user space.
\newblock {\em Communications of the ACM}, 63(6):46--56, 2020.

\bibitem{loukidesswoyer2020}
Mike Loukides and Steve Swoyer.
\newblock Microservices adoption in 2020, Jul 2020.

\bibitem{malkiewicz_2021}
Kylee Malkiewicz.
\newblock This year (so far) in buffer overflows - dover microsystems, Mar
  2021.

\bibitem{mambretti2021bypassing}
Andrea Mambretti, Alexandra Sandulescu, Alessandro Sorniotti, William
  Robertson, Engin Kirda, and Anil Kurmus.
\newblock {Bypassing memory safety mechanisms through speculative control flow
  hijacks}.
\newblock In {\em IEEE European Symposium on Security and Privacy (EuroS\&P)},
  pages 633--649, 2021.

\bibitem{mccanne1993bsd}
Steven McCanne and Van Jacobson.
\newblock The bsd packet filter: A new architecture for user-level packet
  capture.
\newblock In {\em USENIX winter}, volume~46, 1993.

\bibitem{plundervolt}
Kit Murdock, David Oswald, Flavio~D Garcia, Jo~Van~Bulck, Daniel Gruss, and
  Frank Piessens.
\newblock Plundervolt: Software-based fault injection attacks against intel
  sgx.
\newblock In {\em 2020 IEEE Symposium on Security and Privacy (SP)}, pages
  1466--1482. IEEE, 2020.

\bibitem{nagarakatte2009softbound}
Santosh Nagarakatte, Jianzhou Zhao, Milo~MK Martin, and Steve Zdancewic.
\newblock {SoftBound: Highly Compatible and Complete Spatial Memory Safety for
  C}.
\newblock In {\em ACM Conference on Programming Language Design and
  Implementation (PLDI)}, pages 245--258, 2009.

\bibitem{nagarakatte2010cets}
Santosh Nagarakatte, Jianzhou Zhao, Milo~MK Martin, and Steve Zdancewic.
\newblock {CETS: Compiler Enforced Temporal Safety for C}.
\newblock In {\em International Symposium on Memory Management (ISMM)}, pages
  31--40, 2010.

\bibitem{abhaya}
Shankara Pailoor, Xinyu Wang, Hovav Shacham, and Isil Dillig.
\newblock Automated policy synthesis for system call sandboxing.
\newblock {\em Proc. ACM Program. Lang.}, 4(OOPSLA), November 2020.

\bibitem{systrace}
Niels Provos.
\newblock Improving host security with system call policies.
\newblock In {\em Proceedings of the 12th Conference on USENIX Security
  Symposium - Volume 12}, SSYM'03, page~18, USA, 2003. USENIX Association.

\bibitem{priv-esc}
Niels Provos, Markus Friedl, and Peter Honeyman.
\newblock Preventing privilege escalation.
\newblock In {\em Proceedings of the 12th Conference on USENIX Security
  Symposium - Volume 12}, SSYM'03, page~16, USA, 2003. USENIX Association.

\bibitem{ren_liu_2016}
Yuxin Ren, Guyue Liu, Vlad Nitu, Wenyuan Shao, Riley Kennedy, Gabriel Parmer,
  Timothy Wood, and Alain Tchana.
\newblock F2c: Enabling fair and fine-grained resource sharing in multi-tenant
  iaas clouds.
\newblock {\em IEEE Transactions on Parallel and Distributed Systems},
  27(9):2589–2602, 2016.

\bibitem{rice2020}
Liz Rice.
\newblock {\em Container security: fundamental technology concepts that protect
  containerized applications}.
\newblock OReilly Media, 2020.

\bibitem{rudd2017address}
Robert Rudd, Richard Skowyra, David Bigelow, Veer Dedhia, Thomas Hobson,
  Stephen Crane, Christopher Liebchen, Per Larsen, Lucas Davi, Michael Franz,
  Ahmad-Reza Sadeghi, and Hamed Okhravi.
\newblock {Address Oblivious Code Reuse: On the Effectiveness of Leakage
  Resilient Diversity}.
\newblock In {\em Network and Distributed System Security Symposium (NDSS)},
  2017.

\bibitem{saltzer1975protection}
Jerome~H. Saltzer and Michael~D. Schroeder.
\newblock {The Protection of Information in Computer Systems}.
\newblock {\em IEEE}, 63(9):1278--1308, 1975.

\bibitem{schuster2015counterfeit}
Felix Schuster, Thomas Tendyck, Christopher Liebchen, Lucas Davi, Ahmad-Reza
  Sadeghi, and Thorsten Holz.
\newblock {Counterfeit Object-oriented Programming: On the Difficulty of
  Preventing Code Reuse Attacks in C++ Applications}.
\newblock In {\em IEEE Symposium on Security and Privacy (S\&P)}, pages
  745--762, 2015.

\bibitem{serebryany2012addresssanitizer}
Konstantin Serebryany, Derek Bruening, Alexander Potapenko, and Dmitriy Vyukov.
\newblock Addresssanitizer: A fast address sanity checker.
\newblock In {\em 2012 $\{$USENIX$\}$ Annual Technical Conference
  ($\{$USENIX$\}$$\{$ATC$\}$ 12)}, pages 309--318, 2012.

\bibitem{shacham2007geometry}
Hovav Shacham.
\newblock {The Geometry of Innocent Flesh on the Bone: Return-into-libc Without
  Function Calls (on the x86)}.
\newblock In {\em ACM Conference on Computer and Communications Security
  (CCS)}, pages 552--561, 2007.

\bibitem{snow2013just}
Kevin~Z Snow, Fabian Monrose, Lucas Davi, Alexandra Dmitrienko, Christopher
  Liebchen, and Ahmad-Reza Sadeghi.
\newblock {Just-In-Time Code Reuse: On the Effectiveness of Fine-Grained
  Address Space Layout Randomization}.
\newblock In {\em IEEE Symposium on Security and Privacy (S\&P)}, pages
  574--588, 2013.

\bibitem{song2019sok}
Dokyung Song, Julian Lettner, Prabhu Rajasekaran, Yeoul Na, Stijn Volckaert,
  Per Larsen, and Michael Franz.
\newblock Sok: sanitizing for security.
\newblock In {\em 2019 IEEE Symposium on Security and Privacy (SP)}, pages
  1275--1295. IEEE, 2019.

\bibitem{szekeres2013sok}
Laszlo Szekeres, Mathias Payer, Tao Wei, and Dawn Song.
\newblock Sok: Eternal war in memory.
\newblock In {\em 2013 IEEE Symposium on Security and Privacy}, pages 48--62.
  IEEE, 2013.

\bibitem{clkscrew}
Adrian Tang, Simha Sethumadhavan, and Salvatore Stolfo.
\newblock $\{$CLKSCREW$\}$: Exposing the perils of $\{$Security-Oblivious$\}$
  energy management.
\newblock In {\em 26th USENIX Security Symposium (USENIX Security 17)}, pages
  1057--1074, 2017.

\bibitem{team2019}
MSRC Team, Jul 2019.

\bibitem{seccomp}
{The Linux Kernel}.
\newblock Seccomp bpf (secure computing with filters).

\bibitem{w3techs-nginx}
{W3 Techs Web Technology Surveys}.
\newblock Usage statistics of nginx.

\bibitem{w3techs-php}
{W3 Techs Web Technology Surveys}.
\newblock Usage statistics of php for websites.

\bibitem{mining-sandboxes}
Zhiyuan Wan, David Lo, Xin Xia, Liang Cai, and Shanping Li.
\newblock Mining sandboxes for linux containers.
\newblock In {\em 2017 IEEE International Conference on Software Testing,
  Verification and Validation (ICST)}, pages 92--102, 2017.

\end{thebibliography}

\appendix
\section{System Call Policy Comparison Table}
\label{sec:syscalltable}

As part of our evaluation of \timeloops, we generated policies for multiple
programs, as described in Section \ref{sec:evaluation-policies}. In this
section, we list all syscalls contained in the generated policies produced by
\timeloops and \verb$sysfilter$ for our Nginx web application and the
ComposePost microservice of the Social Network service benchmark. We
additionally compare these policies to system calls observed when executing the
programs in a container without any system call filtering, as well to the
default Podman system call filter. We additionally list the most recent Linux
kernel CVE that we were able to identify that could be associated with each
system call.

\onecolumn
\rowcolors{2}{lightgray}{}
\setlength\LTleft{-1cm}
\footnotesize
\begin{center}
\begin{longtable}[c]{ | l | c | c c c | c c c | c |}
    \caption{Comparison of system calls allowed by various filtering policies.
    If a system call does not appear in this table, it was not contained in a
    policy generated by \timeloops, \texttt{sysfilter}, or observed when the
    program executed. The CVE column lists one CVE, if any could be found, in
    the Linux kernel where the associated system call was used to trigger the
    vulnerability.} \\
    \toprule
    & & & \textbf{Nginx} & & & \textbf{ComposePost} & & \\
    \textbf{syscall}
    & \textbf{CVE} & \textbf{Baseline} & \textbf{Timeloops} & \textbf{Sysfilter}
    & \textbf{Baseline} & \textbf{Timeloops} & \textbf{Sysfilter} &
    \textbf{Podman} \\
    \midrule
    \endhead
    \verb$accept$   & CVE-2017-8890 & x & x & x & x & x & x & x \\
    \verb$accept4$  &               & x	& x	& x	&  	& 	& 	& x \\
    \verb$access$   &               & x	& x	& x & x	& x	& x	& x \\
    \verb$alarm$    &               &   &   & x &   &   &   & x \\
    \verb$arch_prctl$ &             & x & x & x & x & x &   & x \\
    \verb$bind$     & CVE-2016-10200 &x & x & x & x & x & x & x \\
    \verb$brk$      & CVE-2020-9391 & x & x & x & x & x & x & x \\
    \verb$capset$   &               & x & x & x & x & x &   & x \\
    \verb$chdir$    &               & x & x & x &   &   &   & x \\
    \verb$chmod$    & CVE-2016-7097 &   &   & x &   &   &   & x \\
    \verb$chown$ & CVE-2015-3339 &  &  & x &  &  &  & x \\
    \verb$clock_getres$ &  & x & x & x &  &  & x & x \\
    \verb$clock_gettime$ & CVE-2011-3209 &  & x & x & x & x & x & x \\
    \verb$clock_nanosleep$ & CVE-2009-2767 &  &  & x &  &  & x & x \\
    \verb$clock_settime$ &  &  &  &  &  &  & x & \\
    \verb$clone$ & CVE-2019-11815 & x & x & x & x & x &  & x \\
    \verb$close$ &  & x & x & x & x & x & x & x \\
    \verb$connect$ &  & x & x & x & x & x & x & x \\
    \verb$dup$ & CVE-2016-3750 & x & x & x &  &  & x & x \\
    \verb$dup2$ &  & x & x & x &  &  &  & x \\
    \verb$epoll_create$ & CVE-2011-1083 & x & x & x &  &  &  & x \\
    \verb$epoll_ctl$ & CVE-2013-7446 & x & x & x &  &  &  & x \\
    \verb$epoll_wait$ &  & x & x & x &  &  &  & x \\
    \verb$eventfd2$ &  & x & x & x &  &  &  & x \\
    \verb$execve$ & CVE-2018-14634 & x & x & x & x & x &  & x \\
    \verb$exit$ &  &  &  & x & x & x & x & x \\
    \verb$exit_group$ &  & x & x & x &  &  & x & x \\
    \verb$fadvise64$ &  &  &  & x &  &  &  & x \\
    \verb$fcntl$ & CVE-2016-7118 & x & x & x & x & x & x & x \\
    \verb$fstat$ &  & x & x & x & x & x & x & x \\
    \verb$ftruncate$ & CVE-2018-18281 &  &  & x &  &  &  & x \\
    \verb$futex$ & CVE-2020-14381 & x & x & x & x & x & x & x \\
    \verb$getcwd$ &  & x & x & x &  &  & x & x \\
    \verb$getdents$ & CVE-2011-1593 & x & x & x &  &  & x & x \\
    \verb$getdents64$ &  &  &  & x &  &  &  & x \\
    \verb$getegid$ &  & x & x &  &  &  &  & x \\
    \verb$geteuid$ &  & x & x & x &  &  &  & x \\
    \verb$getgid$ &  & x & x &  &  &  &  & x \\
    \verb$getpeername$ &  & x & x & x &  &  & x & x \\
    \verb$getpid$ &  & x & x & x &  & x & x & x \\
    \verb$getppid$ &  & x & x & x &  &  &  & x \\
    \verb$getrandom$ &  & x & x & x &  &  & x & x \\
    \verb$getrlimit$ &  &  &  &  & x & x &  & x \\
    \verb$getsockname$ & CVE-2021-38208 & x & x & x & x & x & x & x \\
    \verb$getsockopt$ & CVE-2021-20194 & x & x & x &  &  & x & x \\
    \verb$gettid$ &  &  &  & x &  & x & x & x \\
    \verb$gettimeofday$ &  &  &  & x & x & x & x & x \\
    \verb$getuid$ &  & x & x & x &  &  &  & x \\
    \verb$ioctl$ & numerous drivers & x & x & x &  &  & x & x \\
    \verb$kill$ &  &  & x & x &  &  & x & x \\
    \verb$listen$ &  & x & x & x & x & x & x & x \\
    \verb$lseek$ & CVE-2013-3301 & x & x & x &  &  & x & x \\
    \verb$lstat$ &  & x & x & x &  &  & x & x \\
    \verb$madvise$ & CVE-2016-5195 &  & x & x & x & x & x & x \\
    \verb$mkdir$ &  & x & x & x &  &  &  & x \\
    \verb$mmap$ & CVE-2018-7740 & x & x & x & x & x & x & x \\
    \verb$mprotect$ & CVE-2010-4169 & x & x & x & x & x & x & x \\
    \verb$mremap$ & CVE-2020-10757 &  &  & x &  &  & x & x \\
    \verb$munmap$ & CVE-2020-29369 & x & x & x & x & x & x & x \\
    \verb$nanosleep$ &  &  &  &  &  &  & x & x \\
    \verb$newfstatat$ &  &  &  & x &  &  & x & x \\
    \verb$open$ & CVE-2020-8428 &  & x &  & x & x &  & x \\
    \verb$openat$ & CVE-2020-10768 & x & x & x &  &  & x & x \\
    \verb$pause$ &  &  &  &  &  &  & x & x \\
    \verb$pipe$ & CVE-2015-1805 & x & x &  &  &  &  & x \\
    \verb$poll$ &  & x & x & x & x & x & x & x \\
    \verb$prctl$ & CVE-2020-10768 & x & x & x & x & x &  & x \\
    \verb$pread64$ &  & x & x & x &  &  &  & x \\
    \verb$prlimit64$ &  & x & x & x &  &  & x & x \\
    \verb$pselect6$ &  & x & x &  & x & x &  & x \\
    \verb$pwrite64$ &  & x & x & x &  &  &  & x \\
    \verb$pwritev$ &  &  &  & x &  &  &  & x \\
    \verb$read$ &  & x & x & x & x & x & x & x \\
    \verb$readlink$ & CVE-2011-4077 &  & x & x &  & x &  & x \\
    \verb$readv$ & CVE-2008-3535 & x & x & x &  &  & x & x \\
    \verb$recvfrom$ & CVE-2013-1979 & x & x & x & x & x & x & x \\
    \verb$recvmsg$ & CVE-2013-1979 &  &  & x & x & x & x & x \\
    \verb$rename$ & CVE-2016-6198 &  &  & x &  &  &  & x \\
    \verb$restart_syscall$ & CVE-2014-3180 &  &  & x &  &  &  & x \\
    \verb$rmdir$ &  &  &  & x &  &  &  & x \\
    \verb$rt_sigaction$ &  & x & x & x & x & x & x & x \\
    \verb$rt_sigprocmask$ &  & x & x & x & x & x & x & x \\
    \verb$rt_sigreturn$ & CVE-2017-15537 & x & x & x &  &  & x & x \\
    \verb$rt_sigsuspend$ &  & x & x & x &  &  &  & x \\
    \verb$sched_get_priority_max$ &  &  &  & x & x & x & x & x \\
    \verb$sched_get_priority_min$ &  &  &  & x & x & x & x & x \\
    \verb$sched_getaffinity$ &  &  &  &  &  & x &  & x \\
    \verb$sched_getparam$ &  &  &  & x &  &  & x & x \\
    \verb$sched_getscheduler$ &  &  &  & x &  &  & x & x \\
    \verb$sched_setaffinity$ & CVE-2021-26708 &  &  & x &  &  &  & x \\
    \verb$sched_setscheduler$ &  &  &  & x &  &  & x & x \\
    \verb$sched_yield$ &  &  &  & x &  & x & x & x \\
    \verb$select$ &  & x & x & x &  &  &  & x \\
    \verb$sendfile$ &  &  &  & x &  &  &  & x \\
    \verb$sendmmsg$ & CVE-2011-4594 &  &  & x &  &  & x & x \\
    \verb$sendmsg$ & CVE-2017-17712 &  &  & x &  &  &  & x \\
    \verb$sendto$ & CVE-2017-17712 &  &  & x & x & x & x & x \\
    \verb$set_robust_list$ &  & x & x & x & x & x & x & x \\
    \verb$set_tid_address$ &  & x & x & x & x & x & x & x \\
    \verb$setgid$ & CVE-2021-32760 & x & x & x &  &  & x & x \\
    \verb$setgroups$ & CVE-2018-7169 & x & x & x &  &  & x & x \\
    \verb$setitimer$ &  & x & x & x &  &  &  & x \\
    \verb$setpriority$ &  &  &  & x &  &  &  & x \\
    \verb$setregid$ &  &  &  & x &  &  & x & x \\
    \verb$setresgid$ &  & x & x & x & x & x & x & x \\
    \verb$setresuid$ & CVE-2019-18684 & x & x & x & x & x & x & x \\
    \verb$setreuid$ & CVE-2011-3145 &  &  & x &  &  & x & x \\
    \verb$setrlimit$ &  &  &  &  &  & x &  & x \\
    \verb$setsid$ & CVE-2005-0178 & x & x & x &  &  &  & x \\
    \verb$setsockopt$ & CVE-2016-4998 & x & x & x & x & x & x & x \\
    \verb$setuid$ & CVE-2013-6825 & x & x & x &  &  & x & x \\
    \verb$shmat$ & CVE-2017-5669 &  &  & x &  &  &  & x \\
    \verb$shmdt$ &  &  &  & x &  &  &  & x \\
    \verb$shmget$ & CVE-2017-5669 &  &  & x &  &  &  & x \\
    \verb$shutdown$ &  & x & x & x & x & x & x & x \\
    \verb$sigaltstack$ & CVE-2009-2847 &  & x &  &  & x &  & x \\
    \verb$socket$ & CVE-2017-9074 & x & x & x & x & x & x & x \\
    \verb$socketpair$ & CVE-2010-4249 & x & x & x & x & x &  & x \\
    \verb$stat$ &  & x & x & x &  &  & x & x \\
    \verb$statfs$ &  &  &  & x &  &  &  & x \\
    \verb$sysinfo$ &  & x & x & x &  &  & x & x \\
    \verb$tgkill$ & CVE-2013-2141 &  &  & x &  &  & x & x \\
    \verb$time$ &  &  &  & x &  &  & x & x \\
    \verb$times$ &  & x & x &  &  &  &  & x \\
    \verb$umask$ & CVE-2020-35513 &  &  & x &  &  &  & x \\
    \verb$uname$ & CVE-2012-0957 & x & x & x & x & x & x & x \\
    \verb$unlink$ & CVE-2016-6197 &  &  & x &  &  &  & x \\
    \verb$utimes$ &  &  &  & x &  &  &  & x \\
    \verb$vfork$ & CVE-2005-3106 &  &  &  &  &  & x & x \\
    \verb$wait4$ &  & x & x & x &  &  &  & x \\
    \verb$write$ &  & x & x & x & x & x & x & x \\
    \verb$writev$ & CVE-2016-9755 & x & x & x &  &  & x & x \\
    \bottomrule
\end{longtable}
\end{center}

\end{document}